\documentclass[12pt]{article}
\usepackage{times}
\usepackage{bm}
\usepackage{natbib}
\usepackage{amsmath,amssymb,amsfonts, amsbsy, epsfig, subfig,natbib,soul,hyperref}
\usepackage[usenames]{color}
\usepackage{rotating}
\usepackage{multirow,enumerate}
               
\def\mybox#1{ \begin{center} \bf \purple\vskip-1mm
        \hspace{.0\textwidth}\vbox{\hrule\hbox{\vrule\kern6pt
\parbox{.95\textwidth}{\kern6pt#1\vskip6pt}\kern6pt\vrule}\hrule}
        \end{center} \vskip-5mm}

\newtheorem{theorem}{Theorem}
\newtheorem{corollary}{Corollary}
\newtheorem{proposition}{Proposition}
\newtheorem{lemma}{Lemma}
\newtheorem{example}{Example}
\newtheorem{remark}{Remark}
\newtheorem{definition}{Definition}
\newcommand{\beq}{\begin{equation}}
\newcommand{\eeq}{\end{equation}}
\newcommand{\beas}{\begin{eqnarray*}}
\newcommand{\eeas}{\end{eqnarray*}}
\newcommand{\bea}{\begin{eqnarray}}
\newcommand{\eea}{\end{eqnarray}}
\newcommand{\bei}{\begin{itemize}}
\newcommand{\eei}{\end{itemize}}
\newcommand{\ben}{\begin{enumerate}}
\newcommand{\een}{\end{enumerate}}
\newcommand{\bet}{\begin{theorem}}
\newcommand{\eet}{\end{theorem}}
\newcommand{\bel}{\begin{lemma}}
\newcommand{\eel}{\end{lemma}}
\newcommand{\bep}{\begin{proposition}}
\newcommand{\eep}{\end{proposition}}
\newcommand{\bed}{\begin{definition}}
\newcommand{\eed}{\end{definition}}
\newcommand{\bec}{\begin{corollary}}
\newcommand{\eec}{\end{corollary}}
\newcommand{\bex}{\begin{example}}
\newcommand{\eex}{\end{example}}
\newcommand{\qed}{\quad\hbox{\vrule width 4pt height 6pt depth 1.5pt}}

\def\T{{ \mathrm{\scriptscriptstyle T} }}

\def\0{0}
\def\u{u}

\def\U{U}

\def\A{A}

\def\be{\beta}

\def\eps{\epsilon}
\def\X{X}
\def\R{R}
\def\Y{Y}
\def\Z{Z}
\def\S{\Sigma}
\def\O{\Omega}

\def\a{a}
\def\W{W}
\def\N{N}
\def\D{D}
\def\I{I}

\newcommand{\RR}{\mathbb{R}}

\newbox\TempBox \newbox\TempBoxA

\def\pr{{ \text{pr}}} 
\def\ep{{ E}} 
\def\Cov{{ \text{cov}}} 
\def\Var{{ \text{var}}} 

\definecolor{blue}{RGB}{000,000,200}
\definecolor{green}{RGB}{000,150,100}
\definecolor{purple}{RGB}{200,000,180}
\def\purple{\color{purple}}

\newcommand{\argmin}{\mathop{\rm arg\min}}

\def\text#1{\mbox{\rm #1}}

\def\underwiggle 1{
\ifmmode\setbox\TempBox=\hbox{$ 1$}\else\setbox\TempBox=\hbox{
1}\fi \setbox\TempBoxA=\hbox to \wd\TempBox{\hss\char'176\hss}
\rlap{\copy\TempBox}\smash{\lower9pt\hbox{\copy\TempBoxA}} }

\def\cH{\mathcal{H}}

\newcommand{\real}[1]{\mathrm{I \! R} \mathit{^{#1}}}

\begin{document}

\title{ Hypothesis Testing of Matrix Graph Model with Application to Brain Connectivity Analysis}
\author{Yin Xia \\
\normalsize{\textit{Department of Statistics and Operations Research,}} \\
\normalsize{\textit{University of North Carolina at Chapel Hill, Chapel Hill, NC 27514, U.S.A.}} \\
\normalsize{Email: xiayin@email.unc.edu}\\
\normalsize{and} \\
Lexin Li \\
\normalsize{\textit{Division of Biostatistics,}} \\
\normalsize{\textit{University of California at Berkeley, Berkeley, CA 94720, U.S.A.}} \\
\normalsize{Email: lexinli@berkeley.edu}
}
\date{}

\maketitle

\begin{abstract}
Brain connectivity analysis is now at the foreground of neuroscience research. A connectivity network is characterized by a graph, where nodes represent neural elements such as neurons and brain regions, and links represent statistical dependences that are often encoded in terms of partial correlations. Such a graph is inferred from matrix-valued neuroimaging data such as electroencephalography and functional magnetic resonance imaging. There have been a good number of successful proposals for sparse precision matrix estimation under normal or matrix normal distribution; however, this family of solutions do not offer a statistical significance quantification for the estimated links. In this article, we adopt a matrix normal distribution framework and formulate the brain connectivity analysis as a precision matrix hypothesis testing problem. Based on the separable spatial-temporal dependence structure, we develop oracle and data-driven procedures to test the global hypothesis that all spatial locations are conditionally independent, which are shown to be particularly powerful against the sparse alternatives. In addition,  simultaneous tests for identifying conditional dependent spatial locations with false discovery rate control are proposed in both oracle and data-driven settings. Theoretical results show that the data-driven procedures perform asymptotically as well  as the oracle procedures and enjoy certain optimality properties. The empirical finite-sample performance of the proposed tests is studied via simulations, and the new tests are applied on a real electroencephalography data analysis.
\end{abstract}

\noindent{\bf Key Words:} Connectivity analysis; False discovery rate; Gaussian graphical model; Matrix-variate normal distribution; Multiple testing.

\baselineskip=21pt

\section{Introduction}
\label{intro.sec}

In recent years, matrix-valued data are becoming ubiquitous in a wide range of scientific and business applications, including bioinformatics \citep{YinLi2012}, brain imaging analysis \citep{ReissOgden2010,Aston2012,ZhouLi2014}, finance \citep{LengTang2012}, among many others. Accordingly, the matrix normal distribution is becoming increasingly popular in modeling the matrix-variate observations \citep{Zhou2014}. Our motivating example is an electroencephalography (EEG) data, which measures voltage value from electrodes placed at various brain locations over a period of time for a group of alcoholic subjects and normal controls. One scientific goal is to infer the connectivity patterns among those spatial locations. More generally, accurate and informative mapping of the human connectivity network is now at the center stage of neuroscience research. The objective is to infer brain connectivity network, which is commonly characterized as a graph consisting of nodes and links. Here nodes represent neural elements, from micriscopic neurons to macroscopic brain regions, and links represent statistical dependencies between neural components \citep{Johansen-Berg2013}. Partial correlations, reported by a precision matrix, are frequently employed to describe such statistical dependencies \citep{Fornito2013}. This precision matrix, in turn, is to be derived from imaging modalities, such as EEG, magnetoencephalography, and functional magnetic resonance imaging. The data of those imaging modalities are in the common form of a two-dimensional matrix, with one spatial dimension and the other temporal dimension. 

Adopting a matrix normal distribution framework, we formulate the brain connectivity network analysis as a precision matrix inference problem. Specifically, let $\X \in \real{p \times q}$ denote the spatial-temporal matrix data from an image modality, e.g., EEG. It is assumed to follow a matrix normal distribution with the Kronecker product covariance structure,  
\begin{eqnarray*}
\Cov\{\text{vec}(\X)\} = \S_{L} \otimes \S_{T},
\end{eqnarray*}
where the operator vec$(\X)$ stacks the columns of the matrix $\X$ to a vector, $\otimes$ is the Kronecker product, $\S_{L} \in \real{p \times p}$ denotes the covariance matrix of $p$ spatial locations, $\S_{T} \in \real{q \times q}$ denotes the covariance matrix for $q$ time points. Correspondingly, 
\begin{eqnarray*}
\Cov^{-1}\{\text{vec}(\X)\} = \S_{L}^{-1} \otimes \S_{T}^{-1} = \O_{L} \otimes \O_{T},
\end{eqnarray*}
where $\O_{L} \in \real{p \times p}$ is the spatial precision matrix, $\O_{T} \in \real{q \times q}$ is the temporal precision matrix. In brain connectivity analysis, our primary interest is to infer the the connectivity network characterized by the spatial precision matrix $\O_L$. By contrast, the temporal precision matrix $\O_T$ is of little interest here and is to be treated as a nuisance parameter in our analysis. We also make some remarks regarding the assumptions of our adopted framework. First, the matrix normal assumption has been frequently adopted in various applications \citep{YinLi2012, LengTang2012}, and is scientifically plausible in neuroimaging analysis. For instance, the majority standard neuroimaging processing software, such as SPM \citep{Friston2007} and FSL \citep{Smith2004}, adopt a framework that assumes the data are normally distributed per voxel (location) with a noise factor and an autoregressive structure, which shares a similar spirit as the matrix normal formulation. Second, it is commonly assumed that the precision matrix is sparse, which we adopt for our inferential procedure as well. Again, this sparsity assumption is scientifically justifiable, as it is known that brain region connections are energy consuming \citep{Olshausen2004}, and biological units tend to minimize energy-consuming activities \citep{Bullmore2009}. 

In this article, we aim to address the following two hypothesis testing questions. The first is to test if all spatial locations are conditionally independent, namely, we test the null hypothesis
\beq
\label{global.test}
H_0:~ \O_L \text{ is diagonal } \; \text{ versus } \; H_1:~ \O_L \text{ is not diagonal}.
\eeq
The second is to identify those conditionally dependent pairs of locations with false discovery rate and false discovery proportion control; i.e., we simultaneously test
\beq
\label{entry.test}
H_{0,i,j}: \omega_{L,i,j}= 0 \; \mbox{ versus } \; H_{1,i,j}:  \omega_{L,i,j} \neq 0, \;\; \mbox{ for } \; 1\leq i<j\leq p,
\eeq
where $\omega_{L,i,j}$ is the $(i,j)$th element of $\O_L$. 

In the literature, there have been a good number of methods proposed to estimate a sparse precision matrix under normal distribution \citep{Meinshausen2006, Yuan2007, Friedman2008, Yuan2010, Ravikumar2011, cai2011constrained}. There are extensions of this line of work from a single precision matrix to multiple precision matrices \citep{Danaher2014, ZhuYZ2014}, and from a Gaussian distribution to a more flexible class of nonparanormal distribution \citep{LiuH2012}. Extension of sparse precision matrix estimation has also emerged for matrix-valued data under matrix normal distribution \citep{YinLi2012, LengTang2012, Zhou2014}. However, all those methods tackle the \emph{estimation} aspect of the problem and induce a connectivity network from an estimated precision matrix. Only recently, \emph{hypothesis testing} procedures have been developed under Gaussian graphical model. In particular, \cite{liu2013ggm} proposed a testing procedure to recover a network in the one-sample case, whereas \cite{xia2014} proposed a method to test the equality of two precision matrices, so to infer the differential network structures in genomics. Both papers, however, worked with vector normal data instead of matrix normal data. 

We aim at hypothesis testing for the spatial precision matrix, under the matrix normal framework, to induce the connectivity network of brain regions. We separate the spatial and temporal dependence structures, then infer the precision matrix $\O_L$ through inverse regression models by relating its elements to the regression coefficients. Two procedures are considered. One is to assume the temporal covariance matrix is known, and we term the resulting method as an oracle procedure. The other is to use a data-driven approach to estimate and plug in the temporal covariance matrix, and accordingly we term it a data-driven procedure. We construct test statistics based on the covariances between the residuals from the inverse regression models. We first develop a global test for \eqref{global.test}, and show it is particularly powerful against a sparse alternative, then develop a multiple testing procedure for simultaneously testing the hypotheses \eqref{entry.test} with false discovery rate and false discovery proportion control. We study the theoretical properties of the two testing procedures, in both the oracle and data-driven scenarios. We show that the data-driven procedure performs asymptotically as well  as the oracle procedure, and enjoys certain optimality under the regularity conditions. Our numerical analysis also supports such findings. 

Our contributions are multi-fold. First, brain connectivity analysis is now becoming a central goal of neuroscience research \citep{Fornito2013}, and it constantly calls for statistical significance quantification of the inferred connection between neural elements. However, there is a paucity of  systematic hypothesis testing solutions developed for this type of problems in the literature, and our proposal offers a timely response. Second, although various network sparse estimation methods have been successfully applied in neural connectivity analyses, network hypothesis testing is an utterly different problem than estimation. The key of estimation is to seek a bias-variance tradeoff, and many common sparse estimation solutions such as graphical lasso \citep{Friedman2008} and constrained $\ell_1$-minimization for inverse matrix estimation \citep[clime]{cai2011constrained} are biased estimators. Such estimation methods do not produce a direct quantification of statistical significance for the network edges. By contrast, hypothesis testing starts with an nearly unbiased estimator, and produces an explicit significance quantification. Third, among the few network hypothesis testing solutions, \cite{liu2013ggm} and \cite{xia2014} focused on a vector-valued $\X$ following a normal distribution rather than a matrix-valued data. Directly applying their methods to test the spatial conditional dependence, with no regard for the separable structure of $\O_L$ and $\O_T$, is equivalent to assuming the columns of the matrix $\X$ are independent. This is clearly not true as the measurements at different time points can be highly correlated. Thus it is important to separate the spatial-temporal dependence structure before testing. 

The following notations are used throughout this article. For any matrix normal data $\{\X_k\}_{k=1}^n$, with $\X_k\in \real{p \times q}$, $X_{k,i,l}$ denotes the $i$-th spatial location at the $l$-th time point for the $k$-th sample, and $\bar{X}_{i,l}=\frac{1}{n}\sum_{k=1}^nX_{k,i,l}$; $\X_{k,-i,l}$ denotes the column spatial vector with the $i$-th entry removed, and $\bar{\X}_{\cdot,-i,l}=\frac{1}{n}\sum_{k=1}^n\X_{k,-i,l} \in \real{1\times (p-1)}$. For a $p \times 1$ vector $\a$, $\a_{-i}$ denotes the $(p-1)\times 1$ vector by removing the $i$th entry $a_i$ from $\a$. For an $nq \times p$ data matrix  $\A=(\A_1,\ldots,\A_{nq})^{\T}$ , $\A_{\cdot,-i}$ denotes an $nq\times (p-1)$ matrix $(\A^{\T} _{1,-i},\ldots,\A^{\T} _{nq,-i})^{\T}$, $\bar{\A}_{\cdot,-i}=\frac{1}{nq}\sum_{k=1}^n\A_{k,-i} \in \real{1\times (p-1)}$, $\A_{(i)}=(A_{1,i},\ldots,A_{nq,i})^{\T} \in \real{nq\times 1}$,  $\bar{\A}_{(i)}=(\bar{A}_i,\ldots,\bar{A}_i)^{\T} \in \real{nq\times 1}$, where $\bar{A}_i=\frac{1}{nq}\sum_{k=1}^{nq} A_{k,i}$, and $\bar{\A}_{(\cdot,-i)}=(\bar{\A}_{\cdot,-i}^{\T} ,\ldots,\bar{\A}_{\cdot,-i}^{\T} )^{\T} \in \real{nq\times (p-1)}$. 
Furthermore,  for any $\A\in \real{p \times q}$, $\A_{i,-j}$ denotes the $i$th row of $\A$ with its $j$th entry removed, and $\A_{-i,j}$ denotes the $j$th column of $\A$ with its $i$th entry removed. $\A_{-i,-j}$ denotes the submatrix of $\A$ with its $i$th row and $j$th column removed.  If $\A$ is symmetric, then $\lambda_{\max}(\A)$ and $\lambda_{\min}(\A)$ denote the largest and smallest eigenvalues of $\A$. We also use the following norms. For a vector $\a \in \real{p \times 1}$, define its $\ell_{s}$ norm as $|\a|_{s}=(\sum_{i=1}^{p}|a_{i}|^{s})^{1/s}$ for $1\le s \le \infty$. For a matrix $\A \in \real{p \times p}$, the matrix $1$-norm is  the maximum absolute column sum, $\|\A\|_{L_{1}}=\max_{1\leq j\leq p}\sum_{i=1}^{p}|a_{i,j}|$. The matrix element-wise infinity norm is $\|\A\|_{\infty}=\max_{1\leq i,j\leq p}|a_{i,j}|$ and the element-wise $\ell_1$ norm is $\|\A\|_1=\sum_{i=1}^p\sum_{j=1}^p|a_{i,j}|$. Finally, for two  sequences of real numbers $\{a_{n}\}$ and $\{b_{n}\}$, $a_{n} = O(b_{n})$ if there exists a constant $c$ such that $|a_{n}| \leq c|b_{n}|$ holds for all $n$, $a_{n} = o(b_{n})$ if $\lim_{n\rightarrow\infty}a_{n}/b_{n} = 0$, and $a_{n}\asymp b_{n}$ if there are positive constants $c_0$ and $c_1$ such that $c_0\leq a_{n}/b_{n}\leq c_1$ for all $n$.

\section{Methodology}
\label{test.sec}

In this section, we derive test statistics and testing procedures for the global hypothesis (\ref{global.test}), and the entry-wise hypothesis (\ref{entry.test}) with false discovery rate control.

\subsection{Separation of spatial-temporal dependency}
\label{sec.proc}

Let $\{\X_1,...,\X_{n}\}$, each with dimension $p \times q$, denote $n$ i.i.d.\  random samples from a matrix normal distribution. The mean, without loss of generality, is assumed to be zero, and the covariance is of the form $\S = \S_L \otimes \S_T$. Our interest is to infer about $\O_L = \S_L^{-1}$, while treating $\O_T=\S_T^{-1}$ as a nuisance. To separate the spatial and temporal dependence structures, we build hypothesis testing procedures for (\ref{global.test}) and (\ref{entry.test}) based upon the linear transformation of the original samples $\{\X_{k}{\S}_{T}^{-1/2}, k=1,\ldots, n\}$. Specifically, we consider two scenarios. We first treat $\S_T$ as known, and term the resulting testing procedure an oracle procedure. In practice, however, $\S_T$ is rarely known, and as such we plug in an estimator of $\S_T$. When trace$(\S_L)=p$ holds true, the sample covariance matrix $\hat{\S}_{T}=\frac{1}{np}\sum_{k=1}^n\X_k^{\T} \X_k$ is an unbiased estimator of $\S_{T}$. However, when trace$(\S_L)\neq p$, $\hat{\S}_{T}$ is biased and we have $\ep(\hat{\S}_T)=\{$trace$(\S_L)/p\} \S_T$. As we will show in Remark~\ref{trace} in the next section, trace$(\S_L)$ does not affect our proposed test statistics, and thus we can assume without loss of generality that trace$(\S_L)=p$ to simplify the notations. Subsequently, we plug in the sample covariance matrix $\hat{\S}_T$, develop the hypothesis  testing based on the transformed samples, $\{\X_{k}\hat{\S}_{T}^{-1/2}, k=1,\ldots, n\}$, and term it a data-driven procedure. One may also use other estimators of $\S_T$ or $\O_T$, and we will briefly discuss those alternatives in Section \ref{discuss.sec}.

\subsection{Test statistics}
\label{sec.proc}

We first develop test statistics for the two hypotheses in the oracle case. The development of the data-driven statistics is very similar, so we omit the details and will remark clearly the difference between the oracle and data-driven cases. For simplicity, we also use the same set of notations for the two scenarios, and will only differentiate them when we study their respective asymptotic properties in Section~\ref{theory.sec}. 

It is well established that, under the normal distribution, the precision matrix can be described in terms of regression models \citep[Sec 2.5]{anderson1954introduction}. Specifically, letting $\Y_k = \X_{k}{\S}_{T}^{-1/2}$, $k=1,\ldots,n$, denote the transformed samples, we have, 
\begin{eqnarray} \label{regmdl}
Y_{k,i,l} = \Y_{k,-i,l}^{\T}\be_{i}+\epsilon_{k,i,l}, \quad 1\leq i \leq p, 1\leq l \leq q,
\end{eqnarray}
where $\epsilon_{k,i,l}\sim N(0,\sigma_{L,i,i}-\S_{L,i,-i}\S_{L,-i,-i}^{-1}\S_{L,-i,i})$ is independent of $\Y_{k,-i,l}$. The regression coefficient vector $\be_{i}$ and the error term $\epsilon_{k,i}$ satisfy that 
\[
\be_{i} = -\omega_{L,i,i}^{-1}\O_{L,-i,i}, \quad \text{and} \quad
 r_{i,j} = \Cov(\epsilon_{k,i,l},\epsilon_{k,j,l})=\frac{\omega_{L,i,j}}{\omega_{L,i,i}\omega_{L,j,j}}.
\]
As such, the elements $\omega_{L,i,j}$ of $\O_L$ can be represented in terms of $r_{i,j}$. Next, we construct an estimator of $r_{i,j}$ and its bias-corrected version. We then build on this estimator to obtain an estimator of $\omega_{L,i,j}$, upon which we further develop our test statistics. 

A natural estimator of  $r_{i,j}$ is the sample covariance between the residuals $\hat{\epsilon}_{k,i,l}=Y_{k,i,l}-\bar{Y}_{i,l}-(\Y_{k,-i,l}-\bar{\Y}_{\cdot,-i,l})^{\T}\hat{\be}_{i}$,
\[ 
\tilde{r}_{i,j}=\frac{1}{nq}\sum_{k=1}^{n}\sum_{l=1}^q\hat{\epsilon}_{k,i,l}\hat{\epsilon}_{k,j,l},
\]
where $\hat{\be}_i$, $i=1,\ldots,p$, are estimators of $\be_i$ that satisfy Condition (C1) in the oracle case and satisfies Condition (C1$^\prime$) in the data-driven case, and these estimators can be obtained via standard estimation methods such as the Lasso and Dantzig selector, as will be discussed in Section \ref{conditions}. When $i \ne j$, however, $\tilde{r}_{i,j}$ tends to be biased due to the correlation induced by the estimated parameters. We thus consider a bias-corrected estimator of $r_{i,j}$, 
\[
\hat{r}_{i,j}=-(\tilde{r}_{i,j}+\tilde{r}_{i,i}\hat{\beta}_{i,j}+\tilde{r}_{j,j}\hat{\beta}_{j-1,i}), \mbox{ for }1\leq i<j\leq p.
\]
For $i=j$, we let $\hat{r}_{i,i}=\tilde{r}_{i,i}$, which is a nearly unbiased estimator of $r_{i,i}$. An estimator of the entry $\omega_{L,i,j}$ of the spatial precision matrix $\O_L$ can then be constructed as, 
\[
T_{i,j}=\frac{\hat{r}_{i,j}}{\hat{r}_{i,i}\cdot \hat{r}_{j,j}},  \;\; 1\leq i< j\leq p.
\]
To further estimate the variance of $T_{i,j}$,  note that
\beq\label{theta_ij}
\theta_{i,j} = \Var\{\epsilon_{k,i}\epsilon_{k,j}/(r_{i,i}r_{j,j})\}/(nq) = (1+\rho^2_{i,j})/(nqr_{i,i}r_{j,j}), \quad
\eeq
where $\rho_{i,j}^2={\beta_{i,j}^2r_{i,i}}/{r_{j,j}}$. Then $\theta_{i,j}$ can be estimated by
\[
\hat{\theta}_{i,j} = (1+\hat{\beta}_{i,j}^2\hat{r}_{i,i}/\hat{r}_{j,j})/(nq\hat{r}_{i,i}\hat{r}_{j,j}).
\]
Given $\{T_{i,j}, 1\leq i<j\leq p\}$ are heteroscedastic and can possibly have a wide variability, we standardize $T_{i,j}$ by its standard error, which leads to the standardized statistics,
\[
W_{i,j}=\frac{T_{i,j}}{\sqrt{\hat{\theta}_{i,j}}},  \;\; 1\leq i< j\leq p.
\]
In the next section, we test hypotheses (\ref{global.test}) and (\ref{entry.test}) based on $\{W_{i,j}\}_{i,j=1}^{p}$.

\begin{remark}\label{diff}
Construction of the test statistics for the data-driven procedure is almost the same as that for the oracle procedure, except that the oracle procedure starts with transformed sample $\Y_k = \X_{k}{\S}_{T}^{-1/2}$ in \eqref{regmdl}, whereas the data-driven one replaces it with $\Y_k = \X_{k}\hat{\S}_{T}^{-1/2}$. Furthermore, the regression coefficients slightly vary at different time points in the data-driven scenario, and we shall replace  \eqref{regmdl} by $Y_{k,i,l} = \Y_{k,-i,l}^{\T}\be_{i,l}+\epsilon_{k,i,l}$, for $1\leq i \leq p, 1\leq l \leq q$.
\end{remark}

\begin{remark}\label{trace}
When $\S_T$ is unknown, $\ep(\hat{\S}_T)=\{$trace$(\S_L)/p\} \S_T$. If trace$(\S_L)=cp$, with $c\neq 1$, an unbiased estimator of $\S_T$ becomes $\hat{\S}_T/c$. Accordingly, we shall define the transformed data $\Y_k = \sqrt{c} \X_{k}\hat{\S}_{T}^{-1/2}$, for $k=1,\ldots,n$. Then we have the bias-corrected estimator $\hat{r}_{i,j}=c\hat{r}_{i,j}$, which in turn leads to $T_{i,j} = T_{i,j}/c$, and $\hat{\theta}_{i,j}=\hat{\theta}_{i,j}/c^2$. Thus, the standardized statistic $W_{i,j}$ remains the same, as the constant $c$ is cancelled. Therefore, $c$ does not affect our final test statistics, and thus for notational simplicity, we set $c=1$ from the beginning, without loss of any generality.  
\end{remark}

\subsection{Global testing procedure}
\label{sec.global}

We propose the following test statistic for testing the global null hypothesis $H_0: \O_L$ is diagonal,
\[
M_{nq}=\max_{1\leq i< j\leq p}W_{i,j}^2.
\]
Furthermore, we define the global test $\Psi_\alpha$ by
\[
\Psi_{\alpha}= I( M_{nq}\geq q_{\alpha}+4\log p-\log\log p)
\]
where $q_{\alpha}$ is the $1-\alpha$ quantile of  the type I extreme value distribution with the cumulative distribution function $\exp\{{({8\pi})^{-1/2}}e^{-t/2}\}$, i.e.,
\[
q_\alpha = -\log (8\pi) - 2 \log\log(1-\alpha)^{-1}.
\]
The hypothesis $H_0$ is rejected whenever $\Psi_\alpha =1$.

The above test is developed based on the asymptotic properties of $M_{nq}$, which will be studied in detail in Section \ref{theory.oracle}. Intuitively, $\{W_{i,j}\}_{i,j=1}^{p}$ are approximately standard normal variables under the null distribution, and are only weakly dependent under suitable conditions. Thus $M_{nq}$ is the maximum of the squares of $p(p-1)/{2}$ such random variables,  and its value should be close to $2\log \{p(p-1)/{2}\}\approx 4\log p$ under $H_0$. We will later show that, under certain regularity conditions, $M_{nq}-4\log p-\log \log p$ converges to a type I extreme value distribution under  $H_0$.

\subsection{Multiple testing procedure}
\label{fdr.sec}

Next we develop a multiple testing procedure for $H_{0,i,j}: \omega_{L,i,j}= 0$, so to identify spatial locations that are conditionally dependent. The test statistic $W_{i,j}$ defined in Section~\ref{sec.proc} is employed. Since there are $(p^2-p)/2$ simultaneous hypotheses to test, it is important to control the false discovery rate. Let $t$ be the threshold level such that $H_{0,i,j}$ is rejected if $|W_{i,j}|\geq t$. Let $\mathcal{H}_0=\{(i,j): \O_{L,i,j}=0,1\leq i<j\leq p\}$ be the set of true nulls. Denote by $R_{0}(t) = \sum_{(i,j)\in \mathcal{H}_0}I(|W_{i,j}|\geq t)$ the total number of false positives, and by $R(t)= \sum_{1\leq i<j\leq p}I(|W_{i,j}|\geq t)$ the total number of rejections. The false discovery proportion  and false discovery rate are then defined as 
\[
\text{FDP}(t)=\frac{R_{0}(t)}{R(t)\vee 1}, \quad  \text{FDR}(t)=\ep\{\text{FDP}(t)\}.
\]
An ideal choice of $t$ would reject as many true positives as possible while controlling the false discovery rate and false discovery proportion at the pre-specified level $\alpha$. That is, we select
\[
t_0=\inf\left\{0\leq t\leq 2(\log p)^{1/2}: \; \text{FDP}(t)\leq \alpha\right\}.
\]
We shall estimate $\sum_{(i,j)\in \mathcal{H}_0}I\{|W_{i,j}|\geq t\}$ by $2\{1-\Phi(t)\}|\mathcal{H}_0|$, where $\Phi(t)$ is the standard normal cumulative distribution function.  Note that $|\mathcal{H}_0|$ can be estimated by $(p^2-p)/2$ due to the sparsity of $\O_L$. This leads to the following multiple testing procedure.
\begin{enumerate}[Step 1.]
\item
Calculate the test statistics $W_{i,j}$.
\item
For given $0\leq \alpha\leq 1$, calculate
\[
\hat{t}=\inf\left\{0\leq t\leq 2(\log p)^{1/ 2}: \;  \frac{2\{1-\Phi(t)\}(p^2-p)/2}{R(t)\vee 1}\leq \alpha\right\}.
\]
If $\hat{t}$ does not exist, set $\hat{t}=2(\log p)^{1/ 2}$.
\item
For $1\leq i<j\leq p$, reject $H_{0,i,j}$ if and only if $|W_{i,j}|\geq \hat{t}$.
\end{enumerate}

\section{Theory}
\label{theory.sec}
In this section, we analyze the theoretical properties of the global and multiple testing procedures for both the oracle and data-driven scenarios. We show that the data-driven procedures perform asymptotically as well  as the oracle procedures and enjoy certain optimality under the regularity conditions. For separate treatment of the oracle and data-driven procedures, we now distinguish the notations of the two, and add the superscript ``$o$" to denote the statistics and tests of the oracle procedures, e.g., $\hat{\be}_i^o, M_{nq}^o, \Psi_\alpha^o, \hat{t}^o$, and the superscript ``$d$" to denote those of the data-driven procedures, e.g., $\hat{\be}_i^d, M_{nq}^d, \Psi_\alpha^d$, and $\hat{t}^d$. 

\subsection{Regularity conditions}
\label{conditions}

For the oracle procedure, we require the following set of regularity conditions. 
\begin{enumerate}[(C1)]
\item Assume that $\max_{1\leq i\leq p}|\hat{\be}_{i}^{o} - \be_{i}|_1=o_{\rm p}[\{\log \max(p,q,n)\}^{-1}\}]$, and $\max_{1\leq i\leq p}|\hat{\be}_{i}^{o} - \be_{i}|_2=o_{\rm p}\left\{(nq\log p)^{-1/4}\right\}$. 
\item Assume that $\log p=o\{(nq)^{1/5}\}$,  and there are constants $c_0, c_1>0$ such that, $c_0^{-1}\leq \lambda_{\min}(\O_L)\leq \lambda_{\max}(\O_L)\leq c_0$, and $c_1^{-1}\leq \lambda_{\min}(\O_T)\leq \lambda_{\max}(\O_T)\leq c_1$. 
\item Let $\D_{L}$ be the diagonal  of $\O_L$ and let $\R_{L}=\D_L^{-1/2}\O_L\D_{L}^{-1/2}$ with elements $\eta_{L,i,j}, 1\leq i, j \leq p$. Assume that $\max_{1\leq i< j\leq p} |\eta_{L,i,j} | \leq \eta_L < 1$, for some constant $0 < \eta_L < 1$.
\end{enumerate}

For the data-driven procedure, we replace the above condition (C1) with a slightly different one (C1$^\prime$), then introduce a new condition (C4). 
\begin{enumerate}
\item[\text{(C1$^\prime$)}] Assume that $\max_{1\leq i\leq p,1\leq l\leq q}|\hat{\be}_{i}^{d} - \be_{i,l}|_1=o_{\rm p}[\{\log \max(p,q,n)\}^{-1}\}]$, \\and $\max_{1\leq i\leq p,1\leq l\leq q}|\hat{\be}_{i}^{d} - \be_{i,l}|_2=o_{\rm p}\left\{(nq\log p)^{-1/4}\right\}$. 

\item[\text{(C4)}] Define $s_p=\max_{1\leq l\leq q}\max_{1\leq i\leq p}\sum_{j=1}^p\max\{I(\omega_{L,i,j}\neq 0),I(\omega^d_{l,i,j}\neq 0)\}$, where $(\omega^d_{l,i,j})_{p\times p}$ $=\O^d_l=\Cov^{-1}\{(\X_k\hat{\S}_T^{-1/2})_{\cdot,l}\}$. Assume that $\|\O_T\|_{L_1}^2\|\O_L\|_{L_1}^2=$\\
$o\{\min(r_{1,n,p,q}, r_{2,n,p,q})\}$, where $r_{1,n,p,q}=[np/\{s_p^2q^3\log q\log^3 \max(p,q,n)\}]^{1/2}$, and $r_{2,n,p,q}=(np^2/[s_p^2q^7\{\log q\log\max(p,q,n)\}^2\log p])^{1/4}$.
\end{enumerate}

A few remarks are in order. The estimator $\hat{\be}_{i}^{o}$ satisfying (C1) can be easily obtained via standard estimation methods such as Lasso and Dantzig selector. For instance, if one uses the Lasso estimator, then (C1) is satisfied under (C2) and the sparsity condition $\max_{1\leq i\leq p}|\be_i|_0=o[(nq)^{1/2}/\{\log \max(p,q,n)\}^{3/2}]$. Similarly, $\hat{\be}_{i}^{d}$ satisfying (C1$^\prime$) can be obtained by Lasso if  (C4) holds and the data-driven regression coefficients ${\be}_{i,l}$ satisfy the similar sparsity condition. Conditions (C2) and (C3) are regularity conditions that are commonly used in the high-dimensional hypothesis testing setting \citep{cai2013two,liu2013ggm,xia2014}. (C4) is a mild technical condition. If $\O_T$, $\O_L$ and $\O_l^d$ satisfy $\max_{1\leq i\leq q}\sum_{j=1}^qI(\omega_{T,i,j}\neq 0)\leq s$ and $s_p\leq s$, for some constant $s>0$, then the conditions on matrix 1-norms can be relaxed to the conditions only related to $n, p$ and $q$, namely,  $q^3\log q\log^3\max(p,q,n)=o(np)$ and $q^7\{\log q\log\max(p,q,n)\}^2\log p=o(np^2)$.

\subsection{Oracle global testing procedure}
\label{theory.oracle}

We next analyze the limiting null distribution of the oracle global test statistic $M_{nq}^o$ and the power of the corresponding test $\Psi_{\alpha}^o$. We are particularly interested in the power of the test under the alternative when $\O_L$ is sparse, and show that the power is minimax rate optimal.

The following theorem states the asymptotic null distribution for $M_{nq}^o$, and indicates that, under $H_0$, $M_{nq}^o - 4\log p + \log\log p$ converges weakly to a Gumbel random variable with distribution function $\exp\{-({8\pi})^{-1/2}e^{-t/2}\}$. 
 
\begin{theorem}\label{ther1} Assume (C1), (C2) and (C3). Then under $H_{0}$,  for any $t \in \real{}$,
\begin{eqnarray*} 
\pr(M_{nq}^o - 4\log p+\log\log p\leq t )\rightarrow \exp\{-({8\pi})^{-{1/2}}\exp(-{t/ 2})\}, \; \textrm{ as }  nq,p\rightarrow\infty.
\end{eqnarray*}
Under $H_0$, the above convergence is uniform for all $\{\X_k\}_{k=1}^{n}$ satisfying (C1)-(C3).
\end{theorem}

We next study the power of the corresponding test $\Psi_{\alpha}^o$. We define the following class of precision matrices for spatial locations:
\beq
\label{a1}
\mathcal{U}(c)=\Bigg{\{}\O_L: \max_{1\leq i<j\leq p}\frac{|\omega_{L,i,j}|}{{\theta_{i,j}}^{1/2}}\geq c({\log p})^{1/ 2}\Bigg{\}}.
\eeq
This class of matrices include all precision matrices such that there exists one standardized off-diagonal entry having the magnitude exceeding $c(\log p)^{1/2}$. By the definition in (\ref{theta_ij}), $\theta_{i,j}$ is of the order $1/(nq)$, and thus we only require one of the off-diagonal entries to have size larger than $C \{\log p/(nq)\}^{1/2}$ for some constant $C>0$, where $C$ is fully determined by $c_0$ and $c_1$ in Condition (C2). Then if we choose the constant $c=4$, that is, if there exists one standardized off-diagonal entry having the magnitude larger or equal than $4(\log p)^{1/2}$, the next theorem shows that the null parameter set in which {$\O_L$ is diagonal} is asymptotically distinguishable from $\mathcal{U}(4)$ by the test $\Psi_\alpha^o$. That is,  $H_{0}$ is  rejected by the test $\Psi_{\alpha}^o$ with overwhelming probability if $\O_L \in \mathcal{U}(4)$.

\begin{theorem}\label{th3} 
Assume (C1) and (C2). Then,
\[
\inf_{\O_L\in\mathcal{U}(4)}\pr(\Psi_{\alpha}^o=1)\rightarrow 1, \; \textrm{ as } nq,p \rightarrow\infty.
\]
\end{theorem}

\noindent
The next theorem further shows that this lower bound $4(\log p)^{1/2}$ is rate-optimal. Let ${\cal T}_\alpha$ be the set of   all $\alpha$-level tests, i.e., $\pr(T_{\alpha}=1)\leq \alpha$ under $H_{0}$ for all $T_{\alpha}\in{\cal T}_\alpha$.

\begin{theorem}\label{th3-o} Suppose that $\log p=o(nq)$. Let $\alpha,\beta>0$
and $\alpha+\beta<1$. Then there exists a constant $c_2>0$ such that for all sufficiently large $nq$ and $p$,
\begin{eqnarray*}
\inf_{\O_L\in\mathcal{U}(c_2)}\sup_{T_{\alpha}\in{\cal T}_\alpha}\pr(T_{\alpha}=1)\leq 1-\beta.
\end{eqnarray*}
\end{theorem}

\noindent
As Theorem \ref{th3-o} indicates, if $c_2$ is sufficiently small, then any $\alpha$ level test is unable to reject the null hypothesis correctly uniformly over $\O_L\in\mathcal{U}(c_2)$ with probability tending to one. So the order $(\log p)^{1/2}$ in the lower bound of $ \max_{1\leq i< j\leq p}\{ |\omega_{L,i,j}| \theta_{i,j} ^{-1/2}\}$ in \eqref{a1} cannot be further improved.

\subsection{Oracle multiple testing procedure}

We next investigate the properties of the oracle multiple testing procedure. The following theorem shows that the oracle procedure controls the false discovery proportion and false discovery rate at the pre-specified level $\alpha$ asymptotically.
\bet\label{FDR} Assume (C1) and (C2), and let
\[
\mathcal{S}_\rho= \left\{(i,j): 1\leq i<j\leq p, \frac{|\omega_{L,i,j}|}{\theta_{i,j}^{1/2}} \geq (\log p)^{{1/ 2}+\rho}\right\}.
\]
Suppose for some $\rho, \delta>0$, $|\mathcal{S}_\rho| \geq [{1}/\{({8\pi})^{1/2}\alpha\}+\delta]({\log \log p})^{1/2}$. Suppose $l_0= |\mathcal{H}_0|\geq c_0 p^2$ for some $c_0>0$, and $p\leq c(nq)^r$ for some $c, r > 0$. Letting $l=(p^2-p)/2$, then, 
\[
\lim_{(nq,p)\rightarrow \infty}\frac{\text{FDR}(\hat{t}^o)}{\alpha l_0/l}=1,
\quad
\frac{\text{FDP}(\hat{t}^o)}{\alpha l_0/l} \rightarrow 1 
\]
in probability, as $(nq,p)\rightarrow \infty$. 
\eet
We comment that the condition $|\mathcal{S}_\rho| \geq [{1}/\{({8\pi})^{1/2}\alpha\}+\delta]({\log \log p})^{1/2}$ in Theorem \ref{FDR} is mild, because we have $(p^2-p)/2$ hypotheses in total and this condition only requires a few entries of $\O_L$ having standardized magnitude exceeding $\{(\log p)^{{1/ 2}+\rho}/(nq)\}^{1/2}$ for some constant $\rho>0$.

\subsection{Data-driven procedures}
\label{theory.data}

We next turn to data-driven procedures for both the global testing and the multiple testing. We show that they perform as well as the oracle testing procedures asymptotically. 

\begin{theorem}\label{ther.data} Assume (C1$^\prime$) , (C2)-(C4).
\begin{enumerate}[(i)]
\item 
Under $H_{0}$, for any $t \in \real{}$,
\begin{eqnarray*} 
\pr(M_{nq}^{d}-4\log p+\log\log p\leq t )\rightarrow \exp\{-({8\pi})^{-{1/2}}\exp(-{t/ 2})\}, \; \textrm{ as } nq,p\rightarrow\infty.
\end{eqnarray*}
Under $H_0$, the above convergence is uniform for all $\{\X_k\}_{k=1}^{n}$ satisfying (C1$^\prime$), (C2)-(C4).

\item Furthermore,  
$
\inf_{\O_L\in\mathcal{U}(4)}\pr(\Psi_{\alpha}^{d}=1)\rightarrow 1, \; \textrm{ as } nq,p\rightarrow\infty.
$
\end{enumerate}
\end{theorem}
This theorem shows that $M_{nq}^d$ has the same limiting null distribution as the oracle test statistics $M_{nq}^o$, and the power of the corresponding test $\Psi_\alpha^{d}$ performs as well as the oracle test and is thus minimax rate optimal. The same observation applies to Theorem \ref{FDR.data} below, which shows that the data-driven multiple procedure also performs as well as the oracle case, and controls the false discovery proportion and false discovery rate at the pre-specified level $\alpha$ asymptotically.
\bet\label{FDR.data}
Assume (C1$^\prime$) and (C4). Then under the same conditions as in Theorem \ref{FDR}, 
\[
\lim_{(nq,p)\rightarrow \infty}\frac{\text{FDR}(\hat{t}^{d})}{\alpha l_0/l}=1,
\quad
\frac{\text{FDP}(\hat{t}^{d})}{\alpha l_0/l} \rightarrow 1 
\]
in probability, as $(nq,p)\rightarrow \infty$. 
\eet

\section{Simulations}
\label{simulation.sec}

We study in this section the finite-sample performance of the proposed testing procedures. For the global testing of (\ref{global.test}), we measure the size and power of the oracle test $\Psi_{\alpha}^{o}$ and the data-driven version $\Psi_{\alpha}^{d}$, and for the multiple testing of (\ref{entry.test}), we measure the empirical FDR and power. We compare the oracle and data-driven testing procedures, as well as a simple alternative that was developed by \cite{xia2014} under normal instead of matrix normal distribution, which ignores the separable spatial-temporal structure. The temporal covariance matrix $\S_T$ is constructed with elements $\sigma_{T,i,j}=0.4^{|i-j|}$, $1\leq i,j\leq p$. The sample size and the number of time points is set at $n=20$, $q=20$ and $n=50$, $q=30$, respectively, whereas the spatial dimension $p$ varies among $\{50, 200, 400, 800\}$. We have chosen this setting, since our primary interest is on inferring about spatial connectivity networks with different spatial dimensions. We keep the temporal dimension small, since it is a nuisance in our setup, and choose a relatively small sample size to reflect the fact that there is usually only a limited sample size in many neuroimaging studies. 

For each generated dataset below, we use Lasso to estimate $\be_i$ as
\beq\label{lasso}
\hat{\be}_i=\D_{i}^{-{1\over 2}}\argmin_{\u}\Big\{\frac{1}{2nq}\left|(\Y_{\cdot,-i}-\bar{\Y}_{(\cdot,-i)})\D_{i}^{-1/2}\u-(\Y_{(i)}-\bar{\Y}_{(i)})\right|_2^2+\lambda_{n,i}|\u|_1\Big\},
\eeq
where $\Y$ is the $nq\times p$ data matrix by stacking the transformed samples $\{(\Y_{k,\cdot,l}, k=1,\dots, n, l=1,\dots, q\}$, where $\Y_k = \X_{k}{\S}_{T}^{-1/2}$ for the oracle procedure and $\Y_k = \X_{k}\hat{\S}_{T}^{-1/2}$ for the data-driven procedure, $k=1,\ldots,n$, $\D_{i}=\text{diag}(\hat{\S}_{L,-i,-i})$, and $\hat{\S}_L$ is the sample covariance matrix of $\S_L$ with $nq$ transformed samples, and $\lambda_{n,i}=\kappa \{\hat{\S}_{L,i,i}\log p/(nq) \}^{1/2}$.

\subsection{Global testing simulation}

For the global testing, the data $\{\X_1, \dots, \X_n\}$ are generated from a matrix normal distribution with mean zero and precision matrix $\I\otimes \O_T$ under the null. To evaluate the power, let $\U$ be a matrix with eight random nonzero entries. The locations of  four nonzero entries are selected randomly from the upper triangle of $\U$, each with a magnitude generated  randomly and uniformly from the set $[-4\{{\log p/(nq)}\}^{1/2},-2\{{\log p/(nq)}\}^{1/2}] \cup [2\{{\log p/(nq)}\}^{1/2},4\{{\log p/(nq)}\}^{1/2}]$. The other four nonzero entries in the lower triangle are determined by symmetry. We set $\O_L=(\I+\U+\delta\I)/(1+\delta)$, with $\delta=|\lambda_{\min} (\I+\U)|+0.05$, and choose the tuning parameter $\kappa=2$ in (\ref{lasso}). 

The size and power, in percentage, of the global testing are reported in Table~\ref{tb:test}, based on 1000 data replications and the significance level $\alpha_1=0.05$. We see from Table~\ref{tb:test} that the empirical sizes of the proposed oracle and data-driven procedures are well controlled under the significance level $\alpha_1=0.05$. However, for the vector normal based procedure that ignores the spatial-temporal dependence structure, there is a serious size distortion across all settings. The empirical sizes for the new procedures are slightly below the nominal level for high dimensions, due to the correlation among the variables. Similar phenomenon has also been observed and justified in \citet[Proposition 1]{cai2013two}. We also see from the table that the new procedures are powerful in all settings, even though the two spatial precision matrices differ only in eight entries with the magnitude of difference of the order $\{\log p/(nq)\}^{1/2}$. For both the empirical sizes and powers, the data-driven procedure is seen to perform similarly as the oracle procedure.

\begin{table}[hptb]
   \begin{center}
    \begin{tabular}{|c|c@{\hspace{1em}}|c@{\hspace{1em}}c@{\hspace{1em}} c @{\hspace{1.5em}}c@{\hspace{1.5em}}   |}
       \hline
 &method &\multicolumn{1}{|c|}{$p=50$}&\multicolumn{1}{|c|}{$p=200$}&\multicolumn{1}{|c|}{$p=400$}&\multicolumn{1}{|c|}{$p=800$}\\  [0pt]

  \hline
 
        &&\multicolumn{4}{|c|}{Size}\\
       \hline
        \multirow{3}{*}{$n=30,q=20$}
       &oracle  & 3.6 & 3.5 & 2.8 & 2.9   \\[0pt]
       &data-driven & 3.8 & 3.8 & 2.9 & 2.9   \\[0pt]
       &vector normal & 36.4 & 56.7 & 64.7 & 75.3   \\[0pt]
  \hline
 
        &&\multicolumn{4}{|c|}{Size}\\
       \hline
        \multirow{3}{*}{$n=50,q=30$}
       &oracle  & 3.7 & 3.5 & 5.2 & 4.3   \\[0pt]
       &data-driven & 3.5 & 3.5 & 5.1 &  4.1  \\[0pt]
       &vector normal & 39.7 & 64.8 & 73.5 &  88.4  \\[0pt]
       \hline
        &&\multicolumn{4}{|c|}{Power}\\
       \hline
        \multirow{3}{*}{$n=30,q=20$}
       &oracle  & 77.9 & 85.6 & 87.7 &  90.9  \\[0pt]
       &data-driven & 83.1 & 88.1 & 87.8 & 90.7   \\[0pt]
       &vector normal & 86.2 & 94.1 & 95.0 & 99.1   \\[0pt]
      \hline
        \multirow{3}{*}{$n=50,q=30$}
       &oracle  &62.4  & 74.3 & 68.2 & 75.2   \\[0pt]
       &data-driven & 66.1 & 74.7 & 68.3 & 75.5   \\[0pt]
       &vector normal &76.2  & 92.1 & 93.1 & 90.1   \\[0pt]
       \hline
    \end{tabular}
    \caption{Empirical sizes and powers  (\%)  for global testing. Three methods are compared: the proposed oracle and data-driven procedures based on the matrix normal distribution, and the simple alternative based on the vector normal distribution that ignores the spatial-temporal dependency. The results are based on 1000 data replications.}
\label{tb:test}
\end{center}
\end{table}

\subsection{Multiple testing simulation}

For the multiple testing, the data $\{\X_1, \dots, \X_n\}$ are generated from a matrix normal distribution with mean zero and precision matrix $\O_L\otimes \O_T$. Three choices of $\O_L$ are considered:

\begin{enumerate}[Model 1:]
\item $\O_L^{(1)}=(\omega^{(1)}_{L,i,j})$ where $\omega^{(1)}_{L,i,i}=1$, $\omega^{(1)}_{L,i,i+1}=\omega^{(1)}_{L,i+1,i}=0.6$, $\omega^{(1)}_{L,i,i+2}=\omega^{(1)}_{L,i+2,i}=0.3$ and $\omega^{(1)}_{L,i,j}=0$ otherwise. 

\item $\O_L^{*(2)}=(\omega^{*(2)}_{L,i,j})$ where $\omega^{*(2)}_{L,i,j}=\omega^{*(2)}_{L,j,i}=0.5$ for $i=10(k-1)+1$ and $10(k-1)+2\leq j\leq 10(k-1)+10$, $1\leq k\leq p/10$. $\omega^{*(2)}_{L,i,j}=0$ otherwise. $\O_L^{(2)}=(\O_L^{*(2)}+\delta\I)/(1+\delta)$ with $\delta=|\lambda_{\min}(\O_L^{*(2)})|+0.05$.

\item $\O_L^{*(3)}=(\omega^{*(3)}_{L,i,j})$ where $\omega^{*(3)}_{L,i,i}=1$, $\omega^{*(3)}_{L,i,j}= 0.8\times\text{Bernoulli}(1,2/p)$ for $i < j$ and $\omega^{*(3)}_{L,j,i}=\omega^{*(3)}_{L,i,j}$. $\O_L^{(3)}=(\O_L^{*(3)}+\delta\I)/(1+\delta)$ with $\delta=|\lambda_{\min}(\O_L^{*(3)})|+0.05$.
\end{enumerate}

\noindent
We select the tuning parameters $\lambda_{n,i}$ in (\ref{lasso}) in the Lasso estimation adaptively given the data, following the general principle of making $\sum_{(i,j)\in \mathcal{H}_0}I(|W_{i,j}|\geq t)$ and $\{2-2\Phi(t)\}(p^2-p)/2$ close. The steps of parameter tuning are summarized as follows.

\begin{enumerate}[Step 1.]
\item
Let $\lambda_{n,i}=b/20\sqrt{\hat{\S}_{L,i,i}\log p/(nq)}$, for $b=1,\cdots,40$. For each $b$, calculate $\hat{\be}^{(b)}_{i}$, $i=1,\cdots,p$, and construct the corresponding standardized statistics $W_{i,j}^{(b)}$. 
\item 
Choose $\hat{b}$ as the minimizer of   \[
\sum_{s=1}^{10} \left( \frac{\sum_{(i,j)\in \cH}I(|W_{i,j}|^{(b)}\geq \Phi^{-1}[1-s\{1-\Phi(\sqrt{\log p})\}/10])}{s\{1-\Phi(\sqrt{\log p})\}/10\cdot p(p-1)}-1 \right)^2.
\]
\item 
The tuning parameters $\lambda_{n,i}$ are then set as, 
\[
\lambda_{n,i}=\hat{b}/20\sqrt{\hat{\S}_{L,i,i}\log p/(nq)}.
\]
\end{enumerate}
For comparison, we also carry out the alternative procedure that ignores the Kronecker product structure by using the stacked original data samples $\{\X_{k,\cdot,l}, k=1,\dots, n, l=1,\dots, q\}$.

The empirical FDR and the empirical power of FDR control, in percentage, are summarized in Tables~\ref{tb:FDR1} and \ref{tb:FDR2}, based on 100 data replications and the FDR level set at $\alpha_2=0.1$ and $\alpha_3=0.01$, respectively. In particular, the power is calculated as 
\[
\frac{1}{100}\sum_{l=1}^{100}\frac{\sum_{(i,j)\in \mathcal{H}_1}I(|W_{i,j,l}|\geq \hat{t})}{|\mathcal{H}_1|},
\]
where $W_{i,j,l}$ denotes standardized statistic for the $l$-th replication and $\mathcal{H}_1$ denotes the nonzero locations. We  observe from Tables~\ref{tb:FDR1} and \ref{tb:FDR2} a similar pattern as that from Table~\ref{tb:test}. That is, the empirical FDRs of the proposed oracle and data-driven procedures are both close to the significance levels across all settings, whereas the vector normal based procedure ignoring the spatial-temporal dependence structure yields empirical FDRs much larger than the significance levels. We also see from the table that the new procedures achieve a high power, and the data-driven procedure again performs similarly as the oracle procedure.

\begin{table}[hptb]\addtolength{\tabcolsep}{-4pt}
   \begin{center}
    \begin{tabular}{|c@{\hspace{0.4em}}|c@{\hspace{0.4em}}|r @{\hspace{0.4em}}r @{\hspace{0.4em}}r@{\hspace{0.4em}} r@{\hspace{0.5em}}|r @{\hspace{0.5em}}r@{\hspace{0.5em}}r@{\hspace{0.5em}}r@{\hspace{0.5em}}  |r @{\hspace{0.6em}}r @{\hspace{0.6em}}r@{\hspace{0.6em}} r@{\hspace{0.6em}}|}
       \hline
       & &\multicolumn{4}{|c|}{Model 1}&\multicolumn{4}{|c|}{Model 2}&\multicolumn{4}{|c|}{Model 3}\\[4pt]
       \hline
       $\alpha$ &$p$&\multicolumn{1}{|c|}{ 50}&\multicolumn{1}{|c|}{ 200}&\multicolumn{1}{|c|}{ 400}&\multicolumn{1}{|c|}{ 800}&\multicolumn{1}{|c|}{ 50}&\multicolumn{1}{|c|}{ 200}&\multicolumn{1}{|c|}{ 400}&\multicolumn{1}{|c|}{ 800}&\multicolumn{1}{|c|}{ 50}&\multicolumn{1}{|c|}{ 200}&\multicolumn{1}{|c|}{ 400}&\multicolumn{1}{|c|}{ 800}\\  [4pt]

  \hline
  \hline
        && \multicolumn{12}{|c|}{Empirical FDR  (in \%)}\\[4pt]
       \hline
       \hline
             \multirow{3}{*}{0.1}
       &oracle             & 7.4  & 6.7 & 6.4 & 6.0 & 8.9  & 8.5 & 8.2 & 7.8 & 8.9  & 8.4 & 8.0 & 7.8   \\[4pt]
       &data-driven     & 8.0  & 6.9 & 6.5 & 6.0 & 11.4  & 9.9 & 9.0 & 8.2 & 11.4  & 9.3 & 8.4 & 7.9   \\[4pt]
       &vector normal & 18.9  & 23.1 & 22.7 & 20.7 & 24.5  & 34.1 & 36.5 & 36.5 & 26.1  & 32.0 & 33.5 & 33.8   \\[4pt]
              \hline
              \multirow{3}{*}{0.01}
       &oracle             &  0.6 & 0.5 & 0.4 & 0.7 & 0.9  & 0.8 & 0.6 & 0.5 &  0.8 & 0.7 &0.6 & 0.7 \\[4pt]
       &data-driven     & 0.6  & 0.5 & 0.4 & 0.7 & 1.2  & 0.9 & 0.7 & 0.5 & 1.2  & 0.9 & 0.7&  0.7   \\[4pt]
       &vector normal &  2.5 & 3.2 &2.6  & 5.6 &  4.6 & 5.6 & 5.4 &2.2  & 4.4  & 5.3 & 0.5 & 5.3   \\[4pt]
       \hline
       \hline
        &      & \multicolumn{12}{|c|}{Empirical Power (in \%)}\\[4pt]
       \hline
       \hline
       \multirow{3}{*}{0.1}
       &oracle             & 100.0  & 100.0 & 100.0 & 100.0 & 100.0  & 100.0 & 100.0 & 100.0 & 100.0  & 99.9 & 99.9 & 99.7  \\[4pt]
       &data-driven     & 100.0  & 100.0 & 100.0 & 100.0 & 100.0  & 100.0 & 100.0 & 100.0 & 100.0  & 99.9 & 99.9 &  99.7   \\[4pt]
       &vector normal & 100.0  & 99.9 & 99.9 & 99.9 & 100.0  & 100.0 & 100.0 & 100.0 & 100.0  & 99.8  & 99.6 & 99.2   \\[4pt]

       \hline
       \multirow{3}{*}{0.01}
       &oracle             & 99.9  & 99.9 & 99.8 & 100.0 & 100.0  & 100.0 & 100.0 & 99.8 & 100.0  & 99.7 & 99.2 & 98.5  \\[4pt]
       &data-driven     &  99.9 & 99.9 & 99.8 & 100.0 & 99.9  & 99.9 & 100.0 & 99.8 &  100.0 & 99.6 & 99.1 & 98.5   \\[4pt]
       &vector normal & 99.8  & 99.7 & 99.6 & 99.7 &  99.9 & 99.7 & 99.7 & 99.4 &  99.9 &99.0  & 97.9 & 96.7   \\[4pt]
       \hline
           \end{tabular}
    \caption{Empirical FDRs and powers (\%) for multiple testing with $n=20$ and $q=20$. Three methods are compared: the proposed oracle and data-driven procedures based on the matrix normal distribution, and the simple alternative based on the vector normal distribution that ignores the spatial-temporal dependency. The results are based on 100 data replications.}
\label{tb:FDR1}
\end{center}
\end{table} 

\begin{table}[hptb]\addtolength{\tabcolsep}{-4pt}
   \begin{center}
    \begin{tabular}{|c@{\hspace{0.2em}}|c@{\hspace{0.3em}}|r @{\hspace{0.4em}}r @{\hspace{0.4em}}r@{\hspace{0.4em}} r@{\hspace{0.4em}}|r @{\hspace{0.4em}}r@{\hspace{0.4em}}r@{\hspace{0.4em}}r@{\hspace{0.4em}}  |r @{\hspace{0.4em}}r @{\hspace{0.4em}}r@{\hspace{0.4em}} r@{\hspace{0.4em}}|}
       \hline

       & &\multicolumn{4}{|c|}{Model 1}&\multicolumn{4}{|c|}{Model 2}&\multicolumn{4}{|c|}{Model 3}\\[4pt]
       \hline
       $\alpha$ &$p$&\multicolumn{1}{|c|}{ 50}&\multicolumn{1}{|c|}{ 200}&\multicolumn{1}{|c|}{ 400}&\multicolumn{1}{|c|}{ 800}&\multicolumn{1}{|c|}{ 50}&\multicolumn{1}{|c|}{ 200}&\multicolumn{1}{|c|}{ 400}&\multicolumn{1}{|c|}{ 800}&\multicolumn{1}{|c|}{ 50}&\multicolumn{1}{|c|}{ 200}&\multicolumn{1}{|c|}{ 400}&\multicolumn{1}{|c|}{ 800}\\  [4pt]

  \hline
  \hline
        && \multicolumn{12}{|c|}{Empirical FDR  (in \%)}\\[4pt]
       \hline
       \hline
             \multirow{3}{*}{0.1}
       &oracle             & 8.1  & 7.7 & 8.0 & 7.7 & 8.9  & 9.1 & 8.8 & 8.6 &  8.9 & 8.7 & 8.4 & 8.2   \\[4pt]
       &data-driven     & 8.4  & 7.9 & 8.1 & 7.8 & 10.3  & 10.0 & 9.2 & 8.9 & 11.7  & 9.3 & 8.8 & 8.2   \\[4pt]
       &vector normal &  23.3 &29.8  & 33.6 & 36.1 & 28.2  & 39.8 & 44.4 & 49.2 & 29.0  & 37.5 & 42.2 & 45.8   \\[4pt]
              \hline
              \multirow{3}{*}{0.01}
       &oracle             & 0.7  & 0.6 & 0.6 & 0.6 & 1.1  & 0.8 & 0.7 & 0.8 & 0.7  & 0.8 & 0.8 & 0.7   \\[4pt]
       &data-driven     & 0.7  & 0.6 & 0.6 & 0.6 & 1.2  & 0.9 & 0.8 & 0.8 & 0.9  & 0.9 & 0.8 &  0.7  \\[4pt]
       &vector normal & 3.8  & 4.8 & 5.5 & 5.6 & 5.9  & 7.1 & 7.3 & 9.1 &  5.1 & 6.5 & 7.3 &  8.3  \\[4pt]
       \hline
       \hline
        &      & \multicolumn{12}{|c|}{Empirical Power (in \%)}\\[4pt]
       \hline
       \hline
              \multirow{3}{*}{0.1}
       &oracle             & 100.0  & 100.0 & 100.0 & 100.0 & 100.0  & 100.0 & 100.0 & 100.0 & 100.0  & 100.0 & 100.0 & 100.0  \\[4pt]
       &data-driven     & 100.0  & 100.0 & 100.0 & 100.0 & 100.0  & 100.0 & 100.0 & 100.0 & 100.0  & 100.0 & 100.0 & 100.0  \\[4pt]
       &vector normal & 100.0  & 100.0 & 100.0 & 100.0 & 100.0  & 100.0 & 100.0 & 100.0 & 100.0  & 100.0 & 100.0 & 100.0  \\[4pt]

       \hline
       \multirow{3}{*}{0.1}
       &oracle             & 100.0  & 100.0 & 100.0 & 100.0 & 100.0  & 100.0 & 100.0 & 100.0 & 100.0  & 100.0 & 100.0 & 100.0  \\[4pt]
       &data-driven     & 100.0  & 100.0 & 100.0 & 100.0 & 100.0  & 100.0 & 100.0 & 100.0 & 100.0  & 100.0 & 100.0 & 100.0  \\[4pt]
       &vector normal & 100.0  & 100.0 & 100.0 & 100.0 & 100.0  & 100.0 & 100.0 & 100.0 & 100.0  & 100.0 & 100.0 & 100.0  \\[4pt]
              \hline
           \end{tabular}
    \caption{Empirical FDRs and powers (\%) for multiple testing with $n=50$ and $q=30$. Three methods are compared: the proposed oracle and data-driven procedures based on the matrix normal distribution, and the simple alternative based on the vector normal distribution that ignores the spatial-temporal dependency. The results are based on 100 data replications.}
\label{tb:FDR2}
\end{center}
\end{table}

\section{Real Data Analysis}
\label{application.sec}

We illustrate our testing method on an electroencephalography (EEG) data. The data was collected in a study examining EEG correlates of genetic predisposition to alcoholism and is available at \textsf{http://kdd.ics.uci.edu/datasets/eeg/eeg.data.html}. It consists of 77 alcoholic individuals and 45 controls, and each subject was fitted with a 61-lead electrode cap and was recorded 256 samples per second for one second. There were in addition a ground and two bipolar deviation electrodes, which are excluded from the analysis. The electrode positions were located at standard sites (Standard Electrode Position Nomenclature, American Electroencephalographic Association 1990), and were organized into frontal, central, parietal, occipital, left temporal and right temporal regions. Each subject performed 120 trials under three types of stimuli. More details of data collection can be found in \citet{ZhangEEG1995}. We preprocessed the data in the following ways. Similarly as \citet{LiKimAltman2010}, we focused only on the average of all trials under a single stimulus condition for each subject. We then performed an $\alpha$ band filtering on the signals following \cite{Hayden2006}. Finally, we downsized the temporal dimension by averaging signals at eight consecutive times points. This is to facilitate estimation of the temporal covariance in our testing procedure since it is treated as a nuisance. More discussion of temporal covariance estimation is given in Section~\ref{discuss.sec}. The resulting data is a $61 \times 32$ matrix for each subject, and our goal is to infer the $61 \times 61$ connectivity network of the brain spatial locations.      

\begin{figure}[t]
\centering{
\begin{tabular}{cc}
\includegraphics[height=2.75in,width=2.75in]{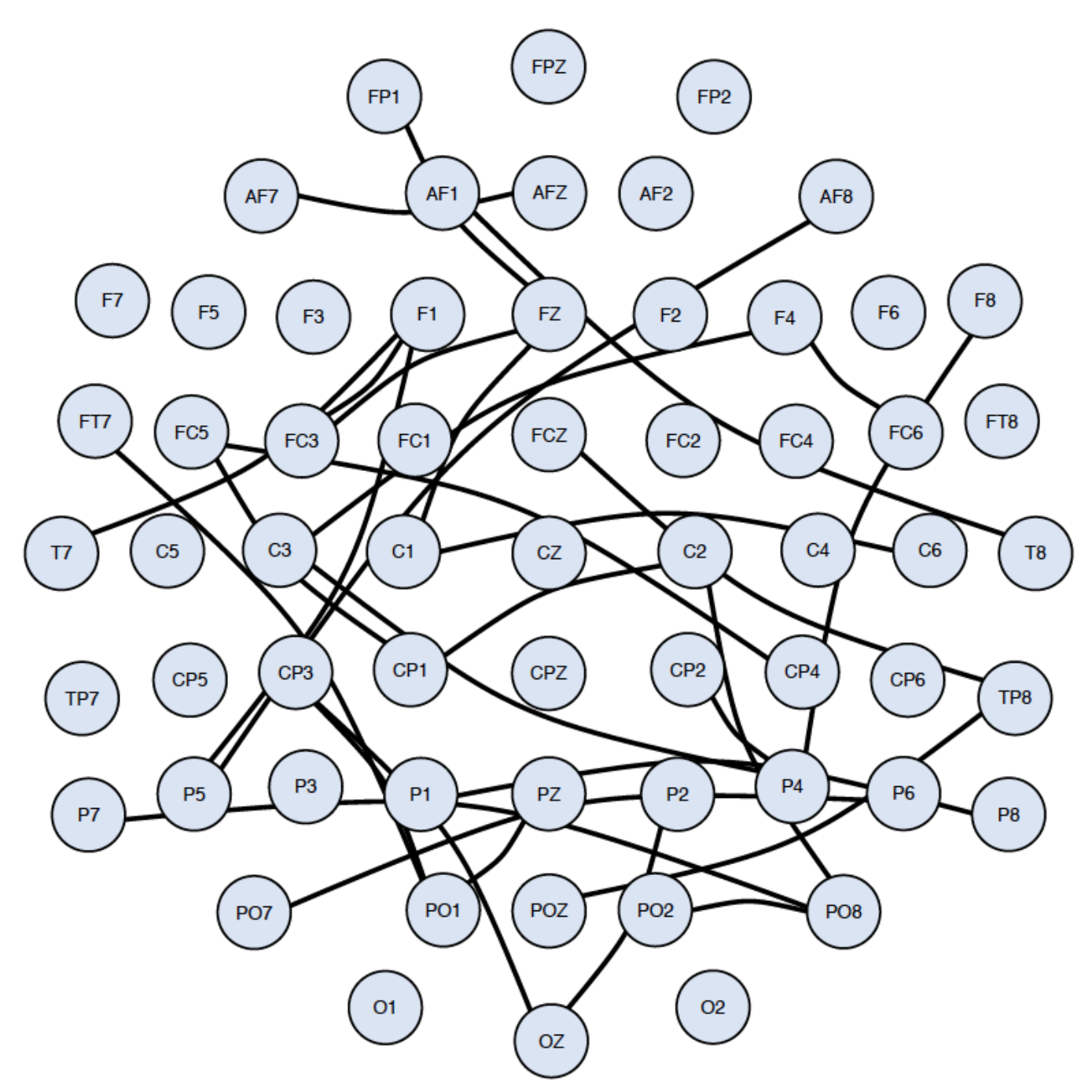} & 
\includegraphics[height=2.75in,width=2.75in]{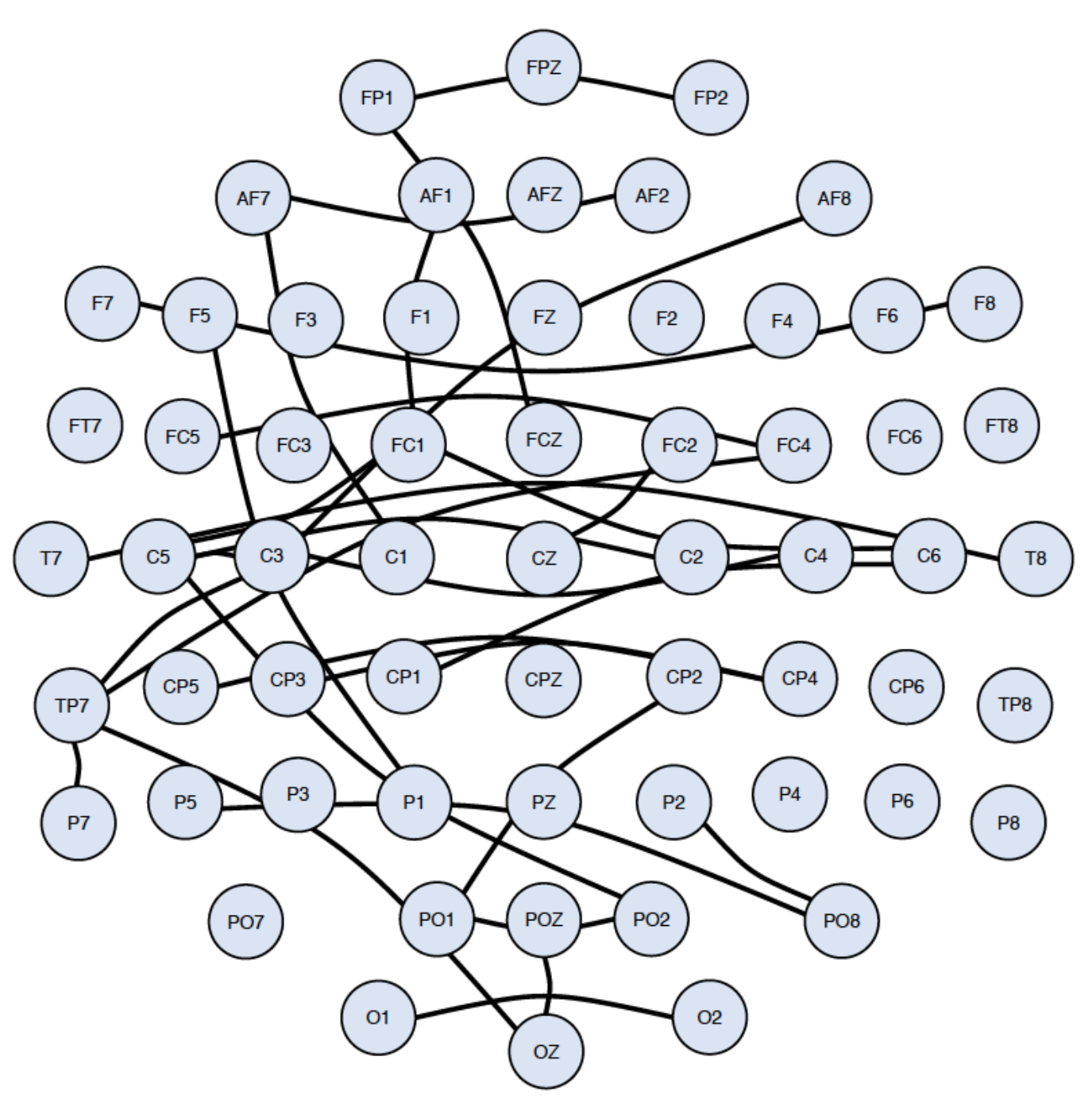} \\
Alcoholic group & Control group \\
\end{tabular}
}
\caption{Connectivity network inferred by the multiple testing procedure for the EEG data. The left panel is for the alcoholic group, and the right panel for the control. Top 30 significant links are shown in this graph.}
\label{fig.eegtest}
\end{figure}

We applied our testing procedures for the alcoholic and control groups separately. We first applied the global test and obtained the $p$-value $0$ for the alcoholic group, and $1.89e^{-15}$ for the control, clearly indicating that some brain regions are connected in the two groups. We then applied the data-driven multiple testing procedure, with a pre-specified FDR significance level $\alpha=0.01$. There were totally $61 \times 60 / 2=1,830$ pairs of spatial locations, among which we identified 208 significant pairs for the alcoholic group, and 155 pairs for the control group. For graphical illustration, we report the top 30 most significant pairs of spatial locations, ordered by their p-values, in Figure~\ref{fig.eegtest}. Examining the connection patterns among those electrodes in the frontal region (denoted by symbols FP, AF, and F), we noted a slight decrease in connections and some asymmetry between the left and right frontal regions in the alcoholic group compared to the control. A similar phenomenon has been observed in \cite{Hayden2006}. We observed similar connection decrease in the central region (FC and C) for the alcoholic group, but connection increase for the parietal region (CP and P). Such findings require additional scientific validation. We also repeated the multiple testing procedure for the downsized EEG data with temporal dimension equal to 16, and observed similar but less clear patterns. For the sake of space, we omit the plots here. In summary, our testing procedure produces a useful list of connections warranted for further examination.

\section{Discussion}
\label{discuss.sec}

We have proposed in this article global and multiple testing procedures under the matrix normal distribution for detecting the conditional dependence between spatial locations. It has been shown that the procedures perform well both theoretically and numerically. In this section, we discuss the strength and some potential limitations of our methods. We also explore some alternatives and point out possible future extension of our work. 

Our work is one of the few to tackle hypothesis testing aspect of connectivity analysis, and is a useful complement to the general literature of statistical estimation and inference regarding high-dimensional precision matrix. The tests and associated theory have been built upon some of the early works on precision matrix inference such as \cite{liu2013ggm} and \cite{xia2014}. However, our work is the first that explicitly exploits the special covariance (precision) structure of a matrix-variate normal distribution, and our simulations suggest using such information would improve the test. A strength of our method is that it works for an adequately large network; in our simulations, the number of nodes varied from 50 to 800, which encompasses the typical region/location-based connectivity analysis in neuroscience. A potential limitation is that our method treats the temporal component as nuisance, and the developed tests hinge on the accuracy of the temporal covariance estimation. In this article, we have simply used the usual sample covariance matrix $\hat{\S}_T$ to estimate $\S_T$. As such the method usually prefers a relatively small temporal dimension. 

Then a potential extension of our proposal is to employ some alternative estimators of $\S_T$. We note that, many such estimators can be used in conjunction with our proposed testing procedures, as long as the estimator $\tilde{\S}_T$ satisfies the condition, 
\[
\|\tilde{\S}_T-c\S_T\|_{\infty}=O_{\rm p}[\{\log q/(np)\}^{1/2}],
\]
for arbitrary constant $c>0$. As an example, if the temporal covariance matrix is sparse, in the sense that $\max_{1\leq i\leq q}\sum_{j=1}^qI(\sigma_{T,i,j}\neq 0)\leq c'$ for some constant $c'>0$, then the adaptive thresholding estimator proposed in \cite{cai2011adaptive} satisfies the above condition and can be used in our testing procedures. Another alternative is that one can directly estimate the precision matrix $\O_T$, and base the testing procedures on $\{\X_k\hat{\O}_T^{1/2}\}_{k=1}^{n}$, as long as the estimator $\tilde{\O}_T$ satisfies that, 
\[
\|{\tilde{\O}}_T-c\O_T\|_{\infty}=O_{\rm p}\Big{[}\|\O_T\|_{L_1}^2\{\log q/(np)\}^{1/2}\Big{]},
\]
for any constant $c>0$. For instance, if the precision matrix $\O_T$ belongs to the following uniformity class of matrices
\[
\mathcal{U}=\Big{\{}\O_T:\max_{1\leq i\leq q}\sum_{j=1}^q|\omega_{T,i,j}|^l\leq s_0(q)\Big{\}},
\]
for $0\leq l<1$, then the clime estimator of \cite{cai2011constrained} can be employed.

In this article, we have primarily focused on the one-sample inference scenario, where we assume the matrix-valued observations came from a single population. In the EEG data analysis, we simply carried out the tests for the two populations separately. Practically, it is of equal interest to study the two-sample scenario, where the aim is to identify changes of the conditional dependence structures for spatial locations across two or more populations. Specifically, let $\X, \Y \in \real{p \times q}$ follow two matrix normal distributions with the Kronecker product covariance structures, $\S\{\text{vec}(\X)\}=\S_{L1} \otimes \S_{T1}, \mbox{  and }\S\{\text{vec}(\Y)\}=\S_{L2} \otimes \S_{T2}$, respectively. The goal becomes the global testing of
\[
H_0:~ \O_{L1} = \O_{L2} \; \text{ versus } \; H_1:~ \O_{L1}\neq  \O_{L2},
\]
and simultaneous testing of
\[
H_{0,i,j}: \omega_{L1,i,j}= \omega_{L2,i,j} \; \mbox{ versus } \; H_{1,i,j}: \omega_{L1,i,j}\neq \omega_{L2,i,j}, \quad\mbox{ $1\leq i<j\leq p$,}
\]
where $\omega_{L1,i,j}$ is the $(i,j)$th element of $\S_{L1}^{-1}=\O_{L1}=(\omega_{L1,i,j})$, and $\omega_{L2,i,j}$ is the $(i,j)$th element of $\S_{L2}^{-1}=\O_{L2}=(\omega_{L2,i,j})$. In the one-sample scenario, we have used the sample covariance matrix $\hat\S_T$ to estimate temporal covariance matrix, because it  can be turned into an unbiased estimator without affecting the test statistics, as shown in Remark~\ref{trace}. However, in the two-sample scenario,  Remark~\ref{trace} can no longer apply, and thus trace$(\S_{L1})$ and trace$(\S_{L2})$ need to be carefully estimated in order to get good estimators for temporal covariance matrices. Consequently, the two-sample comparison is technically much more challenging, and we leave it as our future research.


\section{Appendix: Proofs}
\label{proof.sec}

\subsection{Technical Lemmas}

We prove the main results in this section. We begin by collecting some technical lemmas.

\begin{lemma}[Bonferroni inequality]\label{le0} Let $B=\cup_{t=1}^{p} B_{t}$. For any $k<[p/2]$, we have
\begin{equation*}
\sum_{t=1}^{2k}(-1)^{t-1}F_{t}\leq\pr(B)\leq
\sum_{t=1}^{2k-1}(-1)^{t-1}F_{t},
\end{equation*}
where $ F_{t}=\sum_{1\leq i_{1}<\cdots<i_{t}\leq
p}\pr(B_{i_{1}}\cap\cdots\cap B_{i_{t}}).$
\end{lemma}

Define $U_{i,j}=\frac{1}{nq}\sum_{k=1}^{n}\sum_{l=1}^q\{\epsilon_{k,i,l}\epsilon_{k,j,l}/(r_{i,i}r_{j,j})-\ep\epsilon_{k,i,l}\epsilon_{k,j,l}/(r_{i,i}r_{j,j})\}$. The follow lemma states the results in the oracle case.
\begin{lemma}\label{eq26}
Suppose that (C1) and (C2) hold. Then we have
\[
\max_{1\leq i\leq p}|\hat{r}_{i,i}-r_{i,i}|=o_{\rm p}[\{{\log p/(nq)}\}^{1/2}],
\]
and
\[
\tilde{r}_{i,j}=\tilde{R}_{i,j}-\tilde{r}_{i,i}(\hat{\beta}_{i,j}-\beta_{i,j})-\tilde{r}_{j,j}(\hat{\beta}_{j-1,i}-\beta_{j-1,i})+o_{\rm p}\{(nq\log p)^{-1/2}\},
\]
for $1\leq i<j\leq p$, where $\tilde{R}_{i,j}$ is the empirical covariance between $\{\epsilon_{k,i,l}, k=1,\dots,n, l=1,\dots,q\}$ and $\{\epsilon_{k,j,l}, k = 1, \dots, n, l=1,\dots,q\}$. Consequently, uniformly in $1\leq i<j\leq p$, 
\[
\hat{r}_{i,j}-(\omega_{L,i,i}\hat{\sigma}_{i,i,\epsilon}+\omega_{L,j,j}\hat{\sigma}_{j,j,\epsilon}-1)r_{i,j}=-U_{i,j}+o_{\rm p}\{(nq\log p)^{-1/2}\}, 
\]
where $(\hat{\sigma}_{i,j,\epsilon})=\frac{1}{nq}\sum_{k=1}^{n}\sum_{l=1}^q(\eps_{k,,l}-\bar{\eps}_l)(\eps_{k,,l}-\bar{\eps}_l)'$,  
$\eps_{k,,l}=(\epsilon_{k,1,l},\dots,\epsilon_{k,p,l})$ and $\bar{\eps}_l=\frac{1}{n}\sum_{k=1}^{n}\eps_{k,,l}$.
\end{lemma}
This lemma is essentially proved in \cite{xia2014} with $nq$ inverse regression models instead.

\subsection{Proof of Theorem \ref{ther1}}
Without loss of generality, throughout this section, we assume that $\omega_{L,i,i}=1$ for $i=1,\ldots,p$.
Let
\[
V_{i,j}=-U_{i,j}/\{\Var(\epsilon_{k,i,1}\epsilon_{k,j,1})/nq\}^{1/2}.
\]
By Lemma \ref{eq26}, we have
\beq\label{p2}
\max_{1\leq i\leq p}|\hat{r}_{i,i}-r_{i,i}|=O_{\rm p}[\{{\log p/(nq)}\}^{1\over 2}],
\eeq
and
\[
\max_{1\leq i\leq p}|\hat{r}_{i,i}-\tilde{R}_{i,i}|=o_{\rm p}\{(nq\log p)^{-1/2}\}.
\]
Note that
\beq\label{p1}
\max_{1\leq i< j\leq p}(\hat{\beta}_{i,j}^2\hat{r}_{i,i}/\hat{r}_{j,j}-\rho_{i,j}^2)=o_{\rm p}(1/\log p),
\eeq
Then by Lemma \ref{eq26},
it is easy to see that, under conditions (C1) and (C2), we have under $H_0$
\[
\max_{1\leq i<j\leq p}||W_{i,j}|-|V_{i,j}||=o_{\rm p}\{(\log p)^{-1/2}\}.
\]
Thus it suffices to prove that
\[
\pr(\max_{1\leq i< j\leq p} V^2_{i,j}-4\log p+\log\log p\leq t )\rightarrow \exp\{-({8\pi})^{-{1/ 2}}\exp(-{t/ 2})\}.
\]
We arrange the indices  $\{1\leq i< j\leq p\}$ in any ordering and set them as $\{(i_{m},j_{m}): m=1,\dots,s\}$ with $s=p(p-1)/2$. Let $\theta_{m}=\Var(\epsilon_{k,i_m,l}\epsilon_{k,j_m,l})$,  and define
$Z_{k,m,l} =\epsilon_{k,i_{m},l}\epsilon_{k,j_{m},l}$ for $1\leq k\leq n$ and $1\leq l\leq q$,
$V_{m} =  {(nq\theta_m)^{-1/2}}\sum_{k=1}^{n}\sum_{l=1}^qZ_{k,m,l}$, and 
$\hat{V}_{m} ={(nq\theta_m)^{-1/2}}\sum_{k=1}^{n}\sum_{l=1}^q\hat{Z}_{k,m,l}$,
 where
$ \hat{Z}_{k,m,l}=Z_{k,m,l}I(|Z_{k,m,l}|\leq \tau_{n})-\ep \{Z_{k,m,l}I(|Z_{k,m,l}|\leq \tau_{n})\}$,
 and $\tau_{n}=32\log (p+nq)$.
  Note that
  $\max_{1\leq i<j\leq p}V^{2}_{i,j}=\max_{1\leq m\leq s}V_{m}^{2}$, and that
  \begin{eqnarray*}
\max_{1\leq m\leq s}(nq)^{-{1/ 2}}\sum_{k=1}^{n}\sum_{l=1}^q&&\ep[|Z_{k,m,l}|I\{|Z_{k,m,l}|\geq 32\log (p+nq)\}]\cr
   &&\leq C(nq)^{1/ 2}\max_{1\leq k\leq n}\max_{1\leq l\leq q}\max_{1\leq m\leq s}\ep[|Z_{k,m,l}|I\{|Z_{k,m,l}|\geq 32\log (p+nq)\}]\cr
  &&\leq C(nq)^{1/ 2}(p+nq)^{-4}\max_{1\leq k\leq n}\max_{1\leq l\leq q}\max_{1\leq m\leq s}\ep[|Z_{k,m,l}|\exp\{|Z_{k,m,l}|/8\}]\cr
  &&\leq C(nq)^{1/ 2}(p+nq)^{-4}.
  \end{eqnarray*}
This yields to 
\[
\pr\Big{\{}\max_{1\leq m\leq s}|V_{m}-\hat{V}_{m}|\geq (\log p)^{-1}\Big{\}}\leq \pr\Big{(}\max_{1\leq m\leq s}\max_{1\leq k\leq n}\max_{1\leq l\leq q}|Z_{k,m,l}|\geq
  \tau_{n}\Big{)}=O(p^{-1}).
\]
Note that
\[
  \Big{|}\max_{1\leq m\leq s}V_{m}^{2}-\max_{1\leq m\leq s}\hat{V}_{m}^{2}\Big{|}\leq 2\max_{1\leq m\leq s}|\hat{V}_{m}|\max_{1\leq m\leq s}|V_{m}-\hat{V}_{m}|+\max_{1\leq m\leq s}|V_{m}-\hat{V}_{m}|^{2},
\]
it suffices to prove that for any $t\in \RR$,  as $nq,p\rightarrow\infty$,
\beq\label{p17}
 \pr\Big{(} \max_{1\leq m\leq s}\hat{V}^{2}_{m}-4\log p+\log\log p\leq t\Big{)}\rightarrow \exp\left\{-({8\pi})^{-{1/ 2}}\exp(-{t/ 2})\right\}.
 \eeq
By  Lemma \ref{le0},  for any  integer $l$ with $0<h<s/2$,
\begin{eqnarray}\label{p9}
\sum_{d=1}^{2h}(-1)^{d-1}\sum_{1\leq m_{1}<\cdots<m_{d}\leq q}
\pr\Bigg{(}\bigcap_{j=1}^{d}F_{m_{j}}\Bigg{)}
&\le& \pr\Big{(}\max_{1\leq m\leq s}\hat{V}^{2}_{m}\geq y_{p}\Big{)}\nonumber \\
&\le& \sum_{d=1}^{2h-1}(-1)^{d-1}\sum_{1\leq m_{1}<\cdots<m_{d}\leq q} \pr\Bigg{(}\bigcap_{j=1}^{d}F_{m_{j}}\Bigg{)}, \quad\quad
\end{eqnarray}
where $y_p=4\log p-\log\log p+t$ and
$
F_{m_{j}}=(\hat{V}^{2}_{m_{j}}\geq y_{p}).
$
Let $\tilde{Z}_{k,m,l}=\hat{Z}_{k,m,l}/(\theta_{m})^{1/2}$ for $m=1,\dots,s$ and
$
\W_{k,l}=(\tilde{Z}_{k,m_{1},l},\ldots,\tilde{Z}_{k,m_{d},l}),
$ for $1\leq k\leq n$ and $1\leq l\leq q$. Define $|\a|_{\min}=\min_{1\leq i\leq d}|a_{i}|$ for any vector $\a\in R^{d}$. Then we have
\[
\pr\Bigg{(}\bigcap_{j=1}^{d}F_{m_{j}}\Bigg{)}=\pr\Bigg{(}\Bigg{|}(nq)^{-{1\over 2}}\sum_{k=1}^{n}\sum_{l=1}^q\W_{k,l}\Bigg{|}_{\min}\geq y^{1\over 2}_{p}\Bigg{)}.
\]
Then it follows from Theorem 1 in Za\"{\i}tsev (1987) that
\begin{eqnarray}\label{p10}
\pr\Bigg{(}\Bigg{|}(nq)^{-1/2}\sum_{k=1}^{n}\sum_{l=1}^q\W_{k,l}\Bigg{|}_{\min}\geq y^{1/2}_{p}\Bigg{)}&\leq& \pr
\Big{\{}|\N_{d}|_{\min}\geq y^{1/ 2}_{p}-\epsilon_{n}(\log p)^{-{1/ 2}}\Big{\}}\cr
& &+ c_{1}d^{5\over 2}\exp\Bigg{\{}-\frac{(nq)^{1/2}\epsilon_{n}}{c_{2}d^{3}\tau_{n}(\log p)^{1/2}}\Bigg{\}},
\end{eqnarray}
where $c_{1}>0$ and $c_{2}>0$ are constants,  $\epsilon_{n}\rightarrow 0$ which will be specified later and
 $\N_{d}=(N_{m_{1}},\ldots,N_{m_{d}})$ is a  normal random vector with $\ep (\N_{d})=0$ and $\Cov(\N_{d})=\Cov(\W_{1,1})$.
 Recall that $d$ is a fixed integer which does not depend on $n,p,q$. Because $\log p=o((nq)^{1/5})$, we can let $\epsilon_{n}\rightarrow 0$ sufficiently slowly that, for any large $M>0$
\beq\label{p11}
c_{1}d^{5/ 2}\exp\Bigg{\{}-\frac{(nq)^{1/2}\epsilon_{n}}{c_{2}d^{3}\tau_{n}(\log p)^{1/2}}\Bigg{\}}=O(p^{-M}).
\eeq 
Combining (\ref{p9}), (\ref{p10}) and (\ref{p11}) we have
\beq\label{pp1}
\pr\Big{(}\max_{1\leq m\leq s}\hat{V}^{2}_{m}\geq y_{p}\Big{)}\leq
 \sum_{d=1}^{2h-1}(-1)^{d-1}\sum_{1\leq m_{1}<\cdots<m_{d}\leq s}\pr
\Big{\{}|\N_{d}|_{\min}\geq y^{1/ 2}_{p}-\epsilon_{n}(\log p)^{-{1/ 2}}\Big{\}}+o(1).
\eeq

Similarly, using Theorem 1 in \cite{zaitsev1987gaussian} again, we can get
\beq\label{pp2}
\pr\Big{(}\max_{1\leq m\leq s}\hat{V}^{2}_{m}\geq y_{p}\Big{)}\geq
 \sum_{d=1}^{2h}(-1)^{d-1}\sum_{1\leq m_{1}<\cdots<m_{d}\leq s}\pr
\Big{\{}|\N_{d}|_{\min}\geq y^{1/ 2}_{p}+\epsilon_{n}(\log p)^{-{1/ 2}}\Big{\}}-o(1).
\eeq

We recall the following lemma, which is shown in the supplementary material of \cite{cai2013two}.
\begin{lemma}\label{le4} For any fixed integer $d\geq 1$ and real number $t\in \RR$,
\beq\label{p16}
\sum_{1\leq m_{1}<\cdots<m_{d}\leq q}\pr
\Big{\{}|\N_{d}|_{\min}\geq y^{1/2}_{p}\pm\epsilon_{n}(\log p)^{-1/2}\Big{\}}={1\over d!}\{({8\pi})^{-{1/2}}\exp(-{t/ 2})\}^d\{1+o(1)\}.
\eeq
\end{lemma}

Then Lemma \ref{le4}, (\ref{pp1}) and (\ref{pp2}) yield that
\beas
\limsup_{nq,p\rightarrow\infty}\pr\Big{(}\max_{1\leq m\leq s}\hat{V}^{2}_{m}\geq y_{p}\Big{)}&\leq&  \sum_{d=1}^{2h}(-1)^{d-1}{1\over d!}\{({8\pi})^{-{1/ 2}}\exp(-{t/ 2})\}^d\\
\liminf_{nq,p\rightarrow\infty}\pr\Big{(}\max_{1\leq m\leq s}\hat{V}^{2}_{m}\geq y_{p}\Big{)}&\geq&  \sum_{d=1}^{2h-1}(-1)^{d-1}{1\over d!}\{({8\pi})^{-{1/ 2}}\exp(-{t/2})\}^d
\eeas
for any positive integer $h$. By letting $h\rightarrow\infty$, we obtain (\ref{p17}) and Theorem \ref{ther1} is proved.\qed

\subsection{Proof of Theorem \ref{th3}}
By Lemma \ref{eq26}, we have
\[
\max_{1\leq i<j\leq p}\Big{|}\frac{T_{i,j}-\{1+o(1)\}\ep T_{i,j}}{\hat{\theta}_{i,j}^{1/2}}-V_{i,j}\Big{|}=o_{\rm p}\{(\log p)^{-1/2}\}.
\]
Let
\[
M_{nq}^{1}=\max_{1\leq i<j\leq p}\frac{T_{i,j}-\{1+o(1)\}\ep T_{i,j}}{\hat{\theta}_{i,j}^{1/2}}.
\]
Then the proof of Theorem \ref{ther1} yields
\[
\pr(M_{nq}^1\leq 4\log p- 2^{-1}\log \log p)\rightarrow 1,
\]
as $nq,p\rightarrow \infty$. By (\ref{p2}), (\ref{p1}) and the fact that
\[
\max_{1\leq i< j\leq p}\omega_{L,i,j}^2/\hat{\theta}_{i,j}=\max_{1\leq i< j\leq p}[\{1+o(1)\}\ep T_{i,j}]^2/\hat{\theta}_{i,j}\leq 2M_{nq}^{1}+2M_{nq},
\]
and
the fact that
\[
\max_{1\leq i< j\leq p} |\omega_{L,i,j}|/\theta_{i,j}^{1/2}\geq 4(\log p)^{1/2},
\]
we have
\[
\pr(M_{nq}\geq q_{\alpha}+4\log p-\log\log p)\rightarrow 1
\]
as $nq,p\rightarrow \infty$. \qed

\subsection{Proof of Theorem \ref{th3-o}}
This theorem is essentially proved in \cite{xia2014}, we skip the proof here.

\subsection{Proof of Theorem \ref{FDR}}
By separation of spatial-temporal dependence structure, we have the following $nq$ inverse regression models
\[
(\X_{k}{\S}_{T}^{-1/2})_{i,l} = \alpha_{i,l} + (\X_{k}{\S}_{T}^{-1/2})_{-i,l}\be_{i}+\epsilon_{k,i,l}, \quad 1\leq k\leq n, 1\leq l\leq q.
\]
Then Theorem \ref{FDR} is proved by applying Theorem 3.1 in \cite{liu2013ggm} in $nq$ regression models. \qed

\subsection{Proof of Theorems \ref{ther.data} and \ref{FDR.data}}
We use the superscript ``$d$'' to denote the corresponding statistics we derived from the the data-driven inverse regression models
\[
Y^{d}_{k,i,l} = (\Y_{k,-i,l}^{d})^{\T}\be^{d}_{i,l}+\epsilon^{d}_{k,i,l}, \quad 1\leq i \leq p, 1\leq l \leq q,
\]
where $\Y^{d}_k = \X_{k}\hat{\S}_{T}^{-1/2}$, for $k=1,\ldots,n$.

By the proofs of Theorem \ref{ther1} and Theorem \ref{FDR}, it suffices to prove that uniformly in $1\leq i\leq j\leq p$, 
\beq
\label{aa1}
\hat{r}_{i,j}^{d}-(\omega_{L,i,i}\hat{\sigma}_{i,i,\epsilon}+\omega_{L,j,j}\hat{\sigma}_{j,j,\epsilon}-1)r_{i,j}=-U_{i,j}+o_{\rm p}((nq\log p)^{-1/2}), 
\eeq
where $\hat{r}_{i,j}^{d}=-(\tilde{r}_{i,j}^{d}+\tilde{r}_{i,i}^{d}\hat{\beta}_{i,j}^d+\tilde{r}_{j,j}^{d}\hat{\beta}_{j-1,i}^d)$ with $\tilde{r}_{i,j}^{d}=1/(nq)\sum_{k=1}^n\sum_{l=1}^q\hat{\epsilon}_{k,i,l}^d\hat{\epsilon}_{k,j,l}^d$ and $\hat{\epsilon}_{k,i,l}^d=Y_k^d-(\Y_{k,-i,l}^{d})^{\T}\hat{\be}^{d}_{i,l}$.
Let $\tilde{\epsilon}_{k,i,l}=\epsilon_{k,i,l}-\bar{\epsilon}_{i,l}$. Then we have
\beas
\hat{\epsilon}_{k,i,l}^{d}\hat{\epsilon}_{k,j,l}^{d}&=&\tilde{\epsilon}_{k,i,l}\tilde{\epsilon}_{k,j,l}-\tilde{\epsilon}_{k,i,l}(\Y_{k,-j,l}-\bar{\Y}_{-j,l})^{\T}(\hat{\be}_j^{d}-\be_j)-\tilde{\epsilon}_{k,j,l}(\Y_{k,-i,l}-\bar{\Y}_{-i,l})^{\T}(\hat{\be}_i^{d}-\be_i)\cr
&&+(\hat{\be}_i^{d}-\be_i)^{\T}(\Y_{k,-i,l}-\bar{\Y}_{-i,l})(\Y_{k,-j,l}-\bar{\Y}_{-j,l})^{\T}(\hat{\be}_j^{d}-\be_j)\cr
&&+\Big{[}\{Y_{k,i,l}^{d}-\bar{Y}_{i,l}^{d}-(\Y_{k,-i,l}^{d}-\bar{\Y}_{-i,l}^{d})^{\T}\hat{\be}_{i}^{d}\}\{Y_{k,j,l}^{d}-\bar{Y}_{j,l}^{d}-(\Y_{k,-j,l}^{d}-\bar{\Y}_{-j,l}^{d})^{\T}\hat{\be}_{j}^{d}\}-\cr
&&\quad \{Y_{k,i,l}-\bar{Y}_{i,l}-(\Y_{k,-i,l}-\bar{\Y}_{-i,l})^{\T}\hat{\be}_{i}^{d}\}\{Y_{k,j,l}-\bar{Y}_{j,l}-(\Y_{k,-j,l}-\bar{\Y}_{-j,l})^{\T}\hat{\be}_{j}^{d}\}\Big{]}.
\eeas
Note that
\[
\hat{\S}_T=\frac{1}{np}\sum_{k=1}^n\X_k^{\T}\X_k=\frac{1}{np}\sum_{k=1}^n\Z_k^{\T}\S_L\Z_k,
\]
where $\Z_k=\S_L^{-1/2}\X_k$. Let $\S_L=U^{\T}\Lambda U$, to be the eigen-decomposition of $\S_L$, then we have
\[
\hat{\S}_T=\frac{1}{np}\sum_{k=1}^n(U\Z_k)^{\T}\Lambda (U\Z_k),
\]
where rows of $U\Z_k$ are independent.
Thus it is easy to show that
\[
\|{\hat{\S}}_T-\S_T\|_{\infty}=O_{\rm p}\Big{[}\{\log q/(np)\}^{1/2}\Big{]}.
\]
Thus, we have 
\[
\|{\hat{\S}}_T^{-1}-\S_T^{-1}\|_{\infty}=O_{\rm p}\Big{[}\|\O_T\|_{L_1}^2\{\log q/(np)\}^{1/2}\Big{]}.
\]
This yields that
\[
\|{\hat{\S}}_T^{-1/2}-\S_T^{-1/2}\|_{\infty}=\|({\hat{\S}}_T^{-1/2}+\S_T^{-1/2})^{-1}({\hat{\S}}_T^{-1}-\S_T^{-1})\|_{\infty}=O_{\rm p}[\{q\log q/(np)\}^{1/2}\|\O_T\|_{L_1}^2],
\] 
which implies, uniformly for $l=1,\ldots,q$, $k=1,\ldots,n$,
\[
|(\X_k\hat{\S}_T^{-1/2})_{\cdot,l}-(\X_k{\S}_T^{-1/2})_{\cdot,l}|_{\infty}=O_{\rm p}[\{q^3\log q\log \max(p,q,n)/(np)\}^{1/2}\|\O_T\|_{L_1}^2].
\]
We shall focus on the event $\{\|{\hat{\S}}_T-\S_T\|_{\infty}=O[\{\log q/(np)\}^{1/2}]\}$ in the following analysis. We have, uniformly for $l=1,\ldots,q$, $k=1,\ldots,n$,
\[
\|\Cov^{-1}[(\X_k\hat{\S}_T^{-1/2})_{\cdot,l}]-\O_L\|_{\infty}=O[\{q^3\log q\log \max(p,q,n)/(np)\}^{1/2}\|\O_T\|_{L_1}^2\|\O_L\|_{L_1}^2].
\]
Thus, by (C4), it is easy to show that 
\[
\max_{1\leq i\leq p,1\leq l\leq q}|{\be}_{i,l}^{d}-\be_{i}|_1=o[\{\log \max(p,q,n)\}^{-1}\}],\text{ and }
\max_{1\leq i\leq p,1\leq l\leq q}|{\be}_{i,l}^{d}-\be_{i}|_2=o\left\{(nq\log p)^{-1/4}\right\}.
\]
Thus we have 
\[
\max_{1\leq i\leq p}|\hat{\be}_{i}^{d}-\be_{i}|_1=o_{\rm p}[\{\log \max(p,q,n)\}^{-1}\}],\text{ and }
\max_{1\leq i\leq p}|\hat{\be}_{i}^{d}-\be_{i}|_2=o_{\rm p}\left\{(nq\log p)^{-1/4}\right\}.
\]
Hence, by the proof of Lemma \ref{eq26}, we have
\beas
&&\frac{1}{nq}\sum_{k=1}^n\sum_{l=1}^q\Big{\{}\tilde{\epsilon}_{k,i,l}\tilde{\epsilon}_{k,j,l}-\tilde{\epsilon}_{k,i,l}(\Y_{k,-j,l}-\bar{\Y}_{-j,l})^{\T}(\hat{\be}_j^{d}-\be_j)-\tilde{\epsilon}_{k,j,l}(\Y_{k,-i,l}-\bar{\Y}_{-i,l})^{\T}(\hat{\be}_i^{d}-\be_i)\cr
&&\quad+(\hat{\be}_i^{d}-\be_i)^{\T}(\Y_{k,-i,l}-\bar{\Y}_{-i,l})(\Y_{k,-j,l}-\bar{\Y}_{-j,l})^{\T}(\hat{\be}_j^{d}-\be_j)\Big{\}}\cr
&&\quad\quad =-\frac{1}{nq}\sum_{k=1}^n\sum_{l=1}^q(\hat{\epsilon}_{k,i,l}^{d})^2\hat{\beta}_{i,j}^{d}-\frac{1}{nq}\sum_{k=1}^n\sum_{l=1}^q(\hat{\epsilon}_{k,j,l}^{d})^2\hat{\beta}_{j-1,i}^{d}-\cr
&&\quad\quad\quad(\omega_{L,i,i}\hat{\sigma}_{i,i,\epsilon}+\omega_{L,j,j}\hat{\sigma}_{j,j,\epsilon}-1)r_{i,j}+U_{i,j}+o_{\rm p}((nq\log p)^{-1/2}),
\eeas
uniformly in $1\leq i\leq j\leq p$. 
Note that
\beas
&&\frac{1}{nq}\sum_{k=1}^n\sum_{l=1}^q\Big{[}\{Y_{k,i,l}^{d}-\bar{Y}_{i,l}^{d}-(\Y_{k,-i,l}^{d}-\bar{\Y}_{-i,l}^{d})^{\T}\hat{\be}_{i}^{d}\}\{Y_{k,j,l}^{d}-\bar{Y}_{j,l}^{d}-(\Y_{k,-j,l}^{d}-\bar{\Y}_{-j,l}^{d})^{\T}\hat{\be}_{j}^{d}\}-\cr
&&\quad \{Y_{k,i,l}-\bar{Y}_{i,l}-(\Y_{k,-i,l}-\bar{\Y}_{-i,l})^{\T}\hat{\be}_{i}^{d}\}\{Y_{k,j,l}-\bar{Y}_{j,l}-(\Y_{k,-j,l}-\bar{\Y}_{-j,l})^{\T}\hat{\be}_{j}^{d}\}\Big{]}\cr
&&\quad\quad=\frac{1}{nq}\sum_{k=1}^n\sum_{l=1}^q\Big{(}\Big{\{}Y_{k,i,l}^{d}-\bar{Y}_{i,l}^{d}-(\Y_{k,-i,l}^{d}-\bar{\Y}_{-i,l}^{d})^{\T}\hat{\be}_{i}^{d}\Big{\}}\Big{[}\{Y_{k,j,l}^{d}-\bar{Y}_{j,l}^{d}-(Y_{k,j,l}-\bar{Y}_{j,l})\}\cr
&&\quad\quad\quad-\{\Y_{k,-j,l}^{d}-\bar{\Y}_{-j,l}^{d}-(\Y_{k,-j,l}-\bar{\Y}_{-j,l})\}^{\T}\hat{\be}_{j}^{d}\Big{]}+\Big{\{}Y_{k,j,l}-\bar{Y}_{j,l}-(\Y_{k,-j,l}-\bar{\Y}_{-j,l})^{\T}\hat{\be}_{j}^{d}\Big{\}}\cr
&&\quad\quad\quad\Big{[}\{Y_{k,i,l}^{d}-\bar{Y}_{i,l}^{d}-(Y_{k,i,l}-\bar{Y}_{i,l})\}-\{\Y_{k,-i,l}^{d}-\bar{\Y}_{-i,l}^{d}-(\Y_{k,-i,l}-\bar{\Y}_{-i,l})\}^{\T}\hat{\be}_{i}^{d}\Big{]}\Big{)}
\eeas
It suffices to show that, uniformly in $1\leq i\leq j\leq p$,
\bea
\label{bb1}
s_{1,i,j}&=&\frac{1}{nq}\sum_{k=1}^n\sum_{l=1}^q\Big{\{}Y_{k,i,l}^{d}-\bar{Y}_{i,l}^{d}-(\Y_{k,-i,l}^{d}-\bar{\Y}_{-i,l}^{d})^{\T}\hat{\be}_{i}^{d}\Big{\}}\Big{[}\{Y_{k,j,l}^{d}-\bar{Y}_{j,l}^{d}-(Y_{k,j,l}-\bar{Y}_{j,l})\}\cr
&&-\{\Y_{k,-j,l}^{d}-\bar{\Y}_{-j,l}^{d}-(\Y_{k,-j,l}-\bar{\Y}_{-j,l})\}^{\T}\hat{\be}_{j}^{d}\Big{]}=o_{\rm p}\{(nq\log p)^{-1/2}\}.
\eea
Recall that $\hat{\epsilon}_{k,i,l}^{d}=Y_{k,i,l}^{d}-\bar{Y}_{i,l}^{d}-(\Y_{k,-i,l}^{d}-\bar{\Y}_{-i,l}^{d})^{\T}\hat{\be}_{i}^{d}$, then we have
\beas
s_{1,i,j}=\frac{1}{nq}\sum_{k=1}^n\sum_{l=1}^q\Big{(}\hat{\epsilon}_{k,i,l}^{d}\Big{[}\{Y_{k,j,l}^{d}-\bar{Y}_{j,l}^{d}-(Y_{k,j,l}-\bar{Y}_{j,l})\}-\sum_{h\neq j}\{Y_{k,h,l}^{d}-\bar{Y}_{h,l}^{d}-(Y_{k,h,l}-\bar{Y}_{h,l})\}\hat{\beta}_{h,j}^{d}\Big{]}\Big{)}.
\eeas
Note that $\hat{\epsilon}_{k,i,l}^{d}={\epsilon}_{k,i,l}^{d}+(\Y_{k,-i,l}^{d}-\bar{\Y}_{-i,l}^{d})^{\T}(\be_{i,l}^{d}-\hat{\be}_{i}^{d})$. Let event $A=(\max_{1\leq i\leq p,1\leq l\leq q}|\hat{\be}_{i}^{d}-\be_{i}^{d}|_1=o[\{\log \max(p,q,n)\}^{-1}])$.
Then we have
\beas
\max_{1\leq i\leq p,1\leq l\leq q}|\ep(\hat{\epsilon}_{k,i,l}^{d}|A)|&\leq& \max_{1\leq i\leq p,1\leq l\leq q}|\ep{\epsilon}_{k,i,l}^{d}|+\max_{1\leq i\leq p,1\leq l\leq q}|\ep\{(\Y_{k,-i,l}^{d}-\bar{\Y}_{-i,l}^{d})^{\T}(\be_{i,l}^{d}-\hat{\be}_{i}^{d})|A\}|\cr
&=&o[\{\log \max(p,q,n)\}^{-1/2}].
\eeas
It is easy to check that, for arbitrarily small $\gamma>0$,
\beas
&&\max_{1\leq i\leq p,1\leq l\leq q,1\leq k\leq n}|\{Y_{k,i,l}-\bar{Y}_{i,l}-(Y_{k,i,l}^{d}-\bar{Y}_{i,l}^{d})\}|\cr
&&\quad=\max_{1\leq i\leq p,1\leq l\leq q,1\leq k\leq n}\Big{|}\Big{\{}\Big{(}\X_k-1/n\sum_{k=1}^n\X_k\Big{)}\S_T^{-1/2}\Big{\}}_{i,l}-\Big{\{}\Big{(}\X_k-1/n\sum_{k=1}^n\X_k\Big{)}\hat{\S}_T^{-1/2}\Big{\}}_{i,l}\Big{|}\cr
&&\quad=o_{\rm p}\Big{[}\{q^3\log q\log \max(p,q,n)\log ^{\gamma}p/(np)\}^{1/2}\|\O_T\|_{L_1}^2\Big{]}.
\eeas
Let event 
\beas
&&B=\Big{\{}\max_{1\leq i\leq p,1\leq l\leq q,1\leq k\leq n}|\{Y_{k,i,l}-\bar{Y}_{i,l}-(Y_{k,i,l}^{d}-\bar{Y}_{i,l}^{d})\}|\cr
&&\quad=o\Big{[}\{q^3\log q\log \max(p,q,n)\log ^{\gamma}p/(np)\}^{1/2}\|\O_T\|_{L_1}^2\Big{]}\Big{\}}.
\eeas
Thus, by the fact that $\max_{1\leq i\leq p,1\leq l\leq q,1\leq k\leq n}|\hat{\epsilon}_{k,i,l}^{d}-\epsilon_{k,i,l}|=o_{\rm p}[\{\log \max(p,q,n)\}^{-1/2}]$, it can be shown that, for arbitrarily small $\gamma>0$,
\beas
&&\pr\Big{(}\max_{1\leq i\leq p}\max_{1\leq h\leq p}\Big{|}\frac{1}{nq}\sum_{k=1}^n\sum_{l=1}^q\hat{\epsilon}_{k,i,l}^{d}\{Y_{k,h,l}^{d}-\bar{Y}_{h,l}^{d}-(Y_{k,h,l}-\bar{Y}_{h,l})\}\Big{|}\cr
&&\quad\quad\geq C\sqrt{\frac{q^2\log q\log \max(p,q,n)\log^{1+\gamma} p}{n^2p}}\|\O_T\|_{L_1}^2\Big{)}\cr
&&\quad \leq \pr\Bigg{(}\max_{1\leq i\leq p}\max_{1\leq h\leq p}\Big{|}\frac{1}{nq}\sum_{k=1}^n\sum_{l=1}^q\hat{\epsilon}_{k,i,l}^{d}\{Y_{k,h,l}^{d}-\bar{Y}_{h,l}^{d}-(Y_{k,h,l}-\bar{Y}_{h,l})\}\Big{|}\cr
&&\quad\quad\quad\geq C\sqrt{\frac{q^2\log q\log \max(p,q,n)\log^{1+\gamma} p}{n^2p}}\|\O_T\|_{L_1}^2, A\cap B\Bigg{)}+\pr(A^c)+\pr(B^c)\cr
&&\quad \leq \pr\Big{(}\max_{1\leq i\leq p}\max_{1\leq h\leq p}\Big{|}\frac{1}{\sqrt{nq}}\sum_{k=1}^n\sum_{l=1}^q\hat{\epsilon}_{k,i,l}^{d}\frac{\{Y_{k,h,l}^{d}-\bar{Y}_{h,l}^{d}-(Y_{k,h,l}-\bar{Y}_{h,l})\}}{\sqrt{{q^3\log q\log \max(p,q,n)\log ^{\gamma}p}/{np}}\|\O_T\|_{L_1}^2}\Big{|}\cr
&&\quad\quad\geq C\sqrt{\log p},A\cap B\Big{)}+o(1)=o(1).
\eeas
This, together with (C4), implies that
\beas
&&\max_{1\leq i\leq j\leq p}\Big{|}\sum_{h\neq j}\frac{1}{nq}\sum_{k=1}^n\sum_{l=1}^q\hat{\epsilon}_{k,i,l}^{d}\{Y_{k,h,l}^{d}-\bar{Y}_{h,l}^{d}-(Y_{k,h,l}-\bar{Y}_{h,l})\}\hat{\beta}_{h,j}^{d}\Big{|}\cr
&&\quad=O_{\rm p}\Big{(}\sqrt{\frac{q^2\log q\log \max(p,q,n)\log^{1+\gamma} p}{n^2p}}\Big{)}\|\O_T\|_{L_1}^2\|\O_L\|_{L_1}=o_{\rm p}\{(nq\log p)^{-1/2}\}.
\eeas
Thus (\ref{bb1}) is proved.
Thus we have
\beas
&&\Bigg{|}\frac{1}{nq}\sum_{k=1}^n\sum_{l=1}^q\Big{[}\{Y_{k,i,l}^{d}-\bar{Y}_{i,l}^{d}-(\Y_{k,-i,l}^{d}-\bar{\Y}_{-i,l}^{d})^{\T}\hat{\be}_{i}^{d}\}\{Y_{k,j,l}^{d}-\bar{Y}_{j,l}^{d}-(\Y_{k,-j,l}^{d}-\bar{\Y}_{-j,l}^{d})^{\T}\hat{\be}_{j}^{d}\}-\cr
&&\quad \{Y_{k,i,l}-\bar{Y}_{i,l}-(\Y_{k,-i,l}-\bar{\Y}_{-i,l})^{\T}\hat{\be}_{i}^{d}\}\{Y_{k,j,l}-\bar{Y}_{j,l}-(\Y_{k,-j,l}-\bar{\Y}_{-j,l})^{\T}\hat{\be}_{j}^{d}\}\Big{]}\Bigg{|}\cr
&&\quad\quad =o_{\rm p}\{(nq\log p)^{-1/2}\}
\eeas
uniformly in $1\leq i\leq j\leq p$. 
Hence equation (\ref{aa1}) is proved and Theorems \ref{ther.data} and \ref{FDR.data} thus follow. \qed

\bibliographystyle{apa}
\bibliography{reference3}

\end{document}